\documentclass[aps,prd,twocolumn,superscriptaddress,nofootinbib]{revtex4-1}
\usepackage[colorlinks,citecolor=blue]{hyperref}
\usepackage{amsmath}
\usepackage{amssymb}
\usepackage{graphicx}
\usepackage{slashed}
\usepackage{xcolor}
\usepackage{array}
\usepackage{multirow}
\usepackage{cancel}
\usepackage{color}
\usepackage{float}
\usepackage{enumitem}
\usepackage{ulem}

\allowdisplaybreaks[4]

\newcommand{\ave}[1]{\langle #1\rangle}
\newcommand{\beq}{\begin{equation}}
\newcommand{\eeq}{\end{equation}}
\newcommand{\bea}{\begin{eqnarray}}
\newcommand{\eea}{\end{eqnarray}}
\newcommand{\met}{\slashed{\rm E}_T}
\newcommand{\nn}{\nonumber}

\begin{document}
\title{Measuring Relic Abundance of Minimal Dark Matter at Hadron Colliders}

\author{Qing-Hong Cao}
\email{qinghongcao@pku.edu.cn}
\affiliation{Department of Physics and State Key Laboratory of Nuclear Physics and Technology, Peking University, Beijing 100871, China}
\affiliation{Collaborative Innovation Center of Quantum Matter, Beijing 100871, China}
\affiliation{Center for High Energy Physics, Peking University, Beijing 100871, China}

\author{Ti Gong}
\email{ttigong@pku.edu.cn}
\affiliation{Department of Physics and State Key Laboratory of Nuclear Physics and Technology, Peking University, Beijing 100871, China}

\author{Ke-Pan Xie}
\email{kpxie@snu.ac.kr}
\affiliation{Center for Theoretical Physics, Department of Physics and Astronomy, Seoul National University, Seoul 08826, Korea}

\author{Zhen Zhang}
\email{zh.zhang@pku.edu.cn}
\affiliation{Center for High Energy Physics, Peking University, Beijing 100871, China}

\begin{abstract}

We consider the special case that the dark matter (DM) candidate is not detected in direct-detection programs when the experimental sensitivity reaches the neutrino flux background. In such circumstance the DM searches at the colliders impose constraints on the DM relic abundance if the DM candidate is a WIMPs type. Specifically, we consider the triplet (quintet and septet) DMs in the framework of minimal DM model and explore the potential of discovering the DM candidate in the mono-jet, mono-photon and vector boson fusion channels at the Large Hadron Collider and future 100~TeV hadron collider. If the DM candidate in such a scenario is discovered at the LHC, then additional DM candidates are needed to explain the observed relic abundance. On the other hand, null results in those DM searching programs at the colliders give rise to lower limits of DM relic abundance. 

\end{abstract}
\maketitle

\section{Introduction}

\begin{figure}
\includegraphics[scale=0.23]{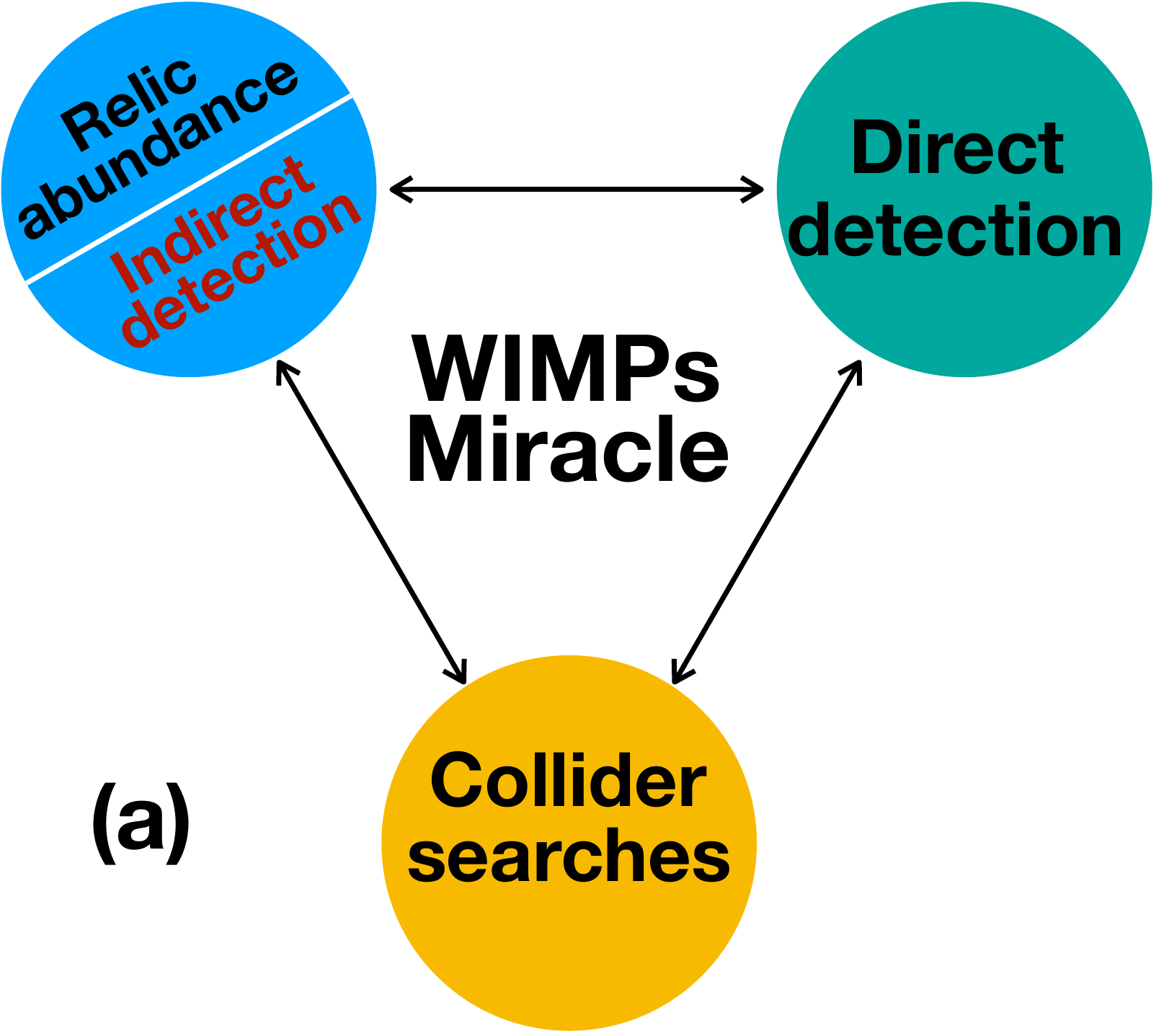}\hspace*{1cm}
\includegraphics[scale=0.23]{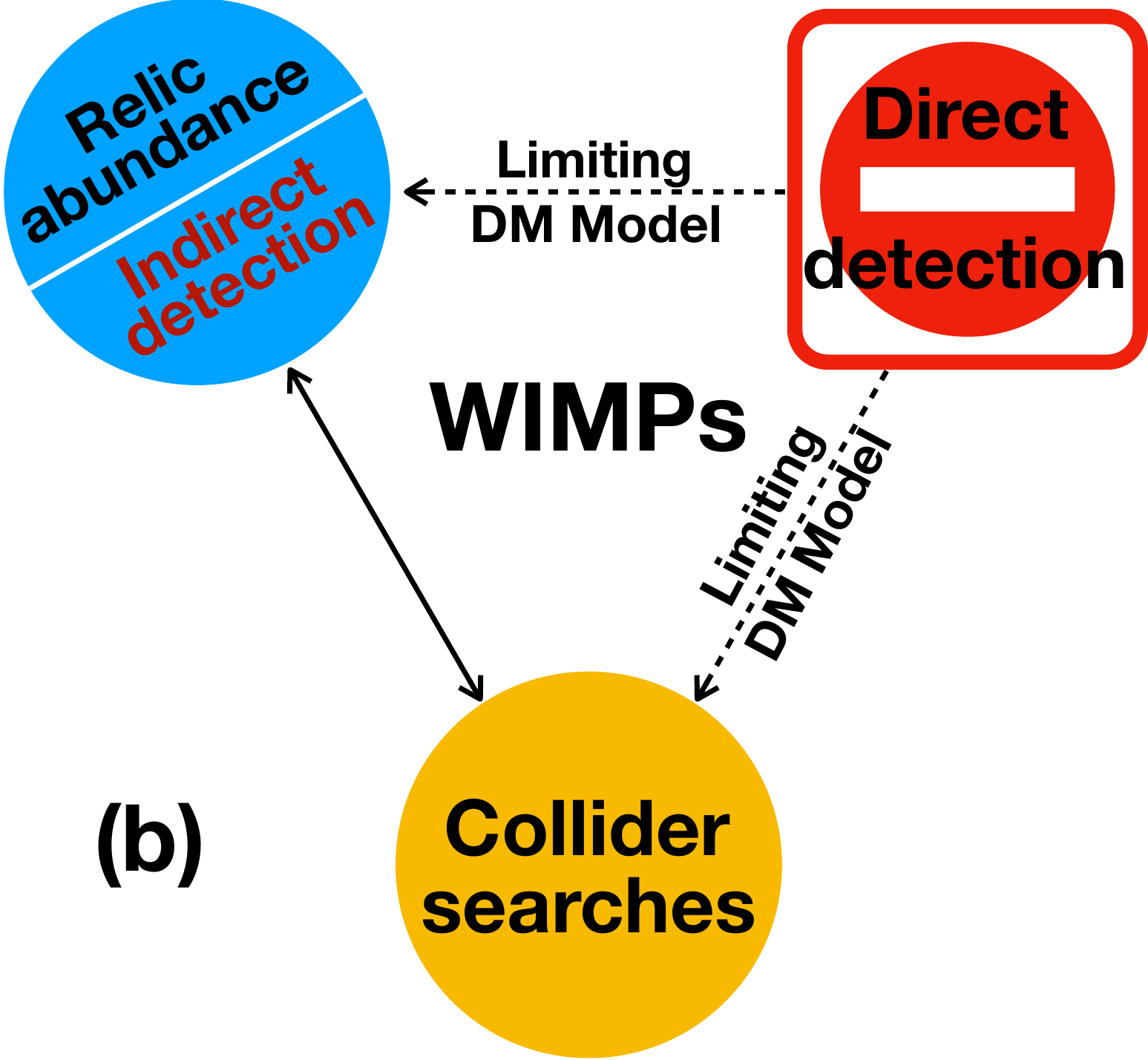}
\caption{Sketch of the impact of null results in the DM direct searches on the WIMPs.  }
\label{fig:wimps}
\end{figure}

The existence of dark matter (DM) is supported by strong evidences including the galaxy rotation curves~\cite{Rubin:1970zza,Rubin:1980zd,Bosma:1981zz,Begeman:1991iy}, the mass distribution of the merging bullet cluster~\cite{Clowe:2006eq}, the cosmic microwave background (CMB)~\cite{Aghanim:2018eyx} and so on. Current cosmological measurements show that the relic abundance of DM is $\Omega_{\rm DM} h^2\approx0.12$, about five times to the visible matter~\cite{Aghanim:2018eyx,Tanabashi:2018oca}. It is known that none of the standard model (SM) particles could serve as a DM candidate~\cite{Bertone:2004pz,Bertone:2010zza,Tanabashi:2018oca}. Among the various new physics (NP) models proposed to explain the particle origin of DM, the class of weakly interacting massive particles (WIMPs) is an attractive and extensively studied scenario. In such a DM paradigm, the DM particles decouple from the thermal plasma of the early Universe at a temperature $T_{\rm F}\sim M_{\rm DM}/20$ ($M_{\rm DM}$ denotes the mass of DM particles), yielding a relic abundance~\cite{Kolb:1990vq,Bertone:2004pz,Bertone:2010zza}
\beq\label{relic}
\Omega_{\rm DM} h^2\sim\frac{0.1\ {\rm pb}\cdot c}{\ave{\sigma v}_{\rm ann}}=0.1\times\frac{3\times10^{-26}~{\rm cm}^3/{\rm s}}{\ave{\sigma v}_{\rm ann}},
\eeq
where $\ave{\sigma v}_{\rm ann}$ denotes the thermal average of the DM annihilation cross section times relative velocity and $c$ is the speed of light. Given the fact that DM particles are non-relativistic when they decouple, $\ave{\sigma v}$ could be expanded in powers of $v^2$ as ~\cite{Kolb:1990vq,Bertone:2004pz,Bertone:2010zza}
\beq\label{annihilation}
\ave{\sigma v}_{\rm ann}=a+b\ave{v^2}+\mathcal{O}(\ave{v^4})\approx a+6b\times \frac{T}{M_{\rm DM}}.
\eeq
Taking $\ave{\sigma v}_{\rm ann}\sim\alpha_{\rm DM}^2/M_{\rm DM}^2$, where $\alpha_{\rm DM}$ stands for the fine structure constant of the interaction between DM and SM particles, we obtain the following estimation
\beq
\Omega_{\rm DM} h^2\sim0.1\left(\frac{0.01}{\alpha_{\rm DM}}\right)^2\left(\frac{M_{\rm DM}}{100~{\rm GeV}}\right)^2.
\eeq
Therefore, if $\alpha_{\rm DM}$ and $M_{\rm DM}$ are of the order of electroweak (EW) coupling and EW scale respectively, we obtain the correct DM relic density. This is called the ``WIMP miracle'', which gives one of the strongest motivations to the WIMPs scenario. More importantly, the WIMPs scenario can be tested from underground direct detection, indirect detection through cosmic rays and collider searches. A global analysis of all kinds of DM searching experiments would help to probe the WIMPs; see the WIMPs miracle triangle loop in Fig.~\ref{fig:wimps}(a).

If the DM candidate is indeed a WIMP, it may be probed by the elastic coherent scattering~\footnote{Note that the coherent enhancement of the elastic scattering cross section is an assumption generally made in the direct detection experiments of DM. Recently, the COHERENT collaboration has observed the elastic neutrino-nucleus coherent scattering which rationalizes the assumption made for DM direct detection to some extent~\cite{Akimov:2017ade}.} on the nucleus (the ``direct detection'')~\cite{Akerib:2016vxi,Aprile:2017iyp,Cui:2017nnn,Jiang:2018pic} or the annihilation to SM particle pairs in the space (the ``indirect detection'')~\cite{Adriani:2008zr,Adriani:2013uda,Aguilar:2013qda,Accardo:2014lma,Aguilar:2014mma}. However, even though the direct detection measurement is becoming more and more accurate and will achieve the irreducible neutrino background in a few years~\cite{Billard:2013qya}, we still have not detected any unquestioned significant signal above the expected backgrounds so far. This null result implies that DM particles may have only tiny interaction with the nucleus, or the detectors are located in a trough of the galactic DM distribution. On the other hand, indirect detection experiments, such as PAMELA and AMS-02, report an ``excess'' of positron fraction $\phi(e^+)/(\phi(e^+)+\phi(e^-))$~\cite{Adriani:2008zr,Aguilar:2013qda,Accardo:2014lma} and positron flux~\cite{Adriani:2013uda,Aguilar:2014mma} in cosmic-ray measurement near the earth, which could be explained by different DM models~\cite{Yin:2008bs,Cao:2007rm,Cirelli:2008pk,Cao:2009yy,Zhang:2009dd,Ibe:2013jya,Jin:2013nta,Feng:2013zca,Ibarra:2013zia,Cao:2014cda}, but also may be explained by other astrophysical sources such as pulsars~\cite{Hooper:2008kg,Linden:2013mqa,Yuan:2013eja,Yin:2013vaa}, or even due to some cosmic-ray propagation effects~\cite{Cowsik:2013woa}. As a result, the collider search may provide a complementary way to investigate the nature of DM.

On the other hand, if the DM candidate is blind to the direct detection experiment, then the WIMPs miracle triangle loop is broken and we can rely only on the interplay between the collider searches and cosmological experiments (relic abundance or the indirect detection). In such a circumstance the direct detection would play a role of DM model killer; see Fig.~\ref{fig:wimps}(b). In this study we focus on the correlation between the collider search  and the relic abundance of DM and demonstrate that it is a no-lose game for searching the DM in the mono-jet channel, the mono-photon channel and the vector-boson-fusion channel; either positive or negative results would shed lights on the DM relic abundance. 

For a given DM model, we could explore the cross section of SM pair generating DM pair at the proton-proton collider, denoted as $\sigma_{\rm pp}$. In general, the $\sigma_{\rm pp}$ correlates with $\sigma_{\rm ann}$ shown in Eq.~(\ref{relic}), despite the fact that the correlation is diluted by some other factors, such as QCD radiative corrections in hadron collisions~\cite{Marchesini:1987cf} and co-annihilation mechanism in the early Universe~\cite{Griest:1990kh}, etc. As the relic abundance $\Omega_{\rm DM}h^2$ is approximately inversely proportional to $\ave{\sigma v}_{\rm ann}$, we are able to transform the collider search constraints to lower limits of DM relic abundance. In this paper, we take the minimal dark matter (MDM) model as an example to demonstrate our approach but we advocate that similar study should be generalized to other WIMP models.

The rest of paper is organized as follows. In Sec.~\ref{sec:models} we briefly introduce the MDM model. In Sec.~\ref{sec:collider} we discuss the collider search phenomenology and its implications on the DM relic abundance. Finally, we summarize in Sec.~\ref{sec:summary}.

\section{The MDM model}\label{sec:models}

In the MDM model, the dark sector consists of an $SU(2)_L$ multiplet with the neutral component being the DM candidate, and no other ingredient is included, which is the meaning of ``minimal''. As a result, given the spin of the dark sector, the only two NP parameters are the weak isospin $j$ and the DM mass $M_{\chi}$. Such a scenario is first proposed by Ref.~\cite{Cirelli:2005uq}, in which the scalar and fermionic MDM models are studied. There have been many studies about the direct and indirect detection signals of MDM, see Refs.~\cite{Cirelli:2009uv} and~\cite{Chua:2013zpa} for the cases of quintet MDM and Dirac fermionic MDM, respectively. Relevant collider studies are also performed for wino- and Higgsino-like MDM in Refs.~\cite{Cirelli:2014dsa,Han:2018wus} at current and future hadron colliders.

In this work we focus on the general fermionic MDM models. In this section we briefly introduce the Lagrangian of Dirac and Majorana fermionic MDM, while the details are discussed in Appendix~\ref{app:MDM}. 

The Dirac MDM with weak isospin $j\geqslant1$ is represented by a multiplet
$$\chi=\begin{pmatrix}\chi_j&\chi_{j-1}&...&\chi_{-j}\end{pmatrix}^T,$$
where $\chi_m$'s are Dirac fermions with the third weak isospin component $T^3=m$. The relevant Lagrangian could be written as
\beq
\label{eq:lagdirac}
\mathcal{L}_\chi^{\rm Dirac}=\bar\chi i\slashed{D}\chi-M_\chi\bar\chi\chi,
\eeq
where
\beq
D_\mu=\partial_\mu-igT^iA_\mu^i-ig'Y_\chi B_\mu,
\eeq
is the gauge covariant derivative with $g$ and $g'$ being couplings for $SU(2)_L$ and $U(1)_Y$ gauge groups, and $T^i$ ($i=1,~2,~3$) are matrix representations of the three generators of the $SU(2)_L$ group. To satisfy the stringent constraint on the spin-independent direct-detection cross section ($\sigma^{\rm SI}$)~\cite{Akerib:2016vxi,Aprile:2017iyp,Cui:2017nnn,Jiang:2018pic}, we set $Y_\chi=0$ to forbid the DM candidate's interaction with $Z$ boson at tree level~\footnote{There are other approaches to release the $\sigma_{\rm SI}$ bound for MDM-like scenarios. For example, a small mass splitting $\Delta m\gtrsim\mathcal{O}(100)$ keV is enough for the Higgsino-like MDM (which has non-zero $Y_\chi$) to satisfy the direct detection bound~\cite{Han:2018wus,Han:1997wn,TuckerSmith:2001hy}.}. Consequently, due to the Gell-Mann-Nishijima relation $Q=T^3+Y_\chi$, $j$ must be an integer to provide an electric neutral component for $\chi$. Expanding Eq.~(\ref{eq:lagdirac}) in components we have
\bea
\label{eq:lag1}
\mathcal{L}_\chi^{\rm Dirac}&=&\sum_{m=-j}^{j}\bar\chi_m(i\slashed{\partial}-M_\chi)\chi_m\nn\\
&+&\sum_{m=-j}^{j}Q_mg(c_WZ_\mu+s_WA_\mu)\bar\chi_m\gamma^\mu\chi_m\nn\\
&+&\sum_{m=-j+1}^{j}\sqrt{(j+m)(j-m+1)}\frac{g}{\sqrt{2}}W^-_\mu\bar\chi_{m-1}\gamma^\mu\chi_m\nn\\
&+&{\rm h.c.},
\eea
where $Q_m\equiv m$ is the electric charge of field $\chi_m$, and $s_W$ and $c_W$ are the sine and cosine of the Weinberg angle. The dark particles' masses are degenerate at the tree-level, but electroweak loop corrections break the mass degeneracy and induce a mass split $\delta M\sim (M_Z-M_W)/(4\pi)^2\sim100$ MeV. The one-loop calculation shows that~\cite{Cirelli:2005uq}
\beq
M_m-M_{m'}\sim 166~{\rm MeV}\times (Q_m^2-Q_{m'}^2),
\eeq
thus the neutral component $\chi_0$ is the lightest one in the dark sector and serves as the DM candidate. The charged component could decay to lighter components plus pions or leptons; for example, for the quintet Majorana fermionic MDM, the charged component with a mass of 4.4 TeV has a 97.7\% branching ratio to $\pi^\pm$ and a lifetime of 1.8 cm~\cite{Cirelli:2005uq}. The neutral component could experience the elastic scattering with the nucleus via loop diagrams~\cite{Cirelli:2005uq}, which are suppressed enough to survive under the current direct detection bounds.

For the Majorana case, we first use the 2-component Weyl spinor with weak isospin $j$
$$\xi=\begin{pmatrix}\xi_j&\xi_{j-1}&...&\xi_{-j}\end{pmatrix}^T,$$
and write down the Lagrangian as
\beq
\mathcal{L}_\xi^{\rm Majorana}=\xi^\dagger i\bar\sigma^\mu D_\mu\xi-\frac{M_\chi}{2}(\xi U \xi+{\rm h.c.}),
\eeq
where the $(2j+1)\times(2j+1)$ matrix $U$ satisfies $L^\dagger UL=U$ with $L$ being the group element of $SU(2)_L$. In this case, $Y_\xi=0$ is also required by the $U(1)_Y$ invariance of the mass term. Translating into 4-component language we have one neutral Majorana DM candidate $\chi_0$, and $2j$ charged Dirac fermions $\chi_m$ and $\bar\chi_m$ with $m=1,~2,~\cdots,~j$, and the Lagrangian is written as
%\begin{widetext}
\begin{align}
\mathcal{L}_\chi^{\rm Majorana}=&\frac{1}{2}\chi_0(i\slashed{\partial}-M_\chi)\chi_0+\sum_{m=1}^{j}\bar\chi_m(i\slashed{\partial}-M_\chi)\chi_m\nn\\
+&\sum_{m=1}^{j}Q_mg(c_WZ_\mu+s_WA_\mu)\bar\chi_m\gamma^\mu\chi_m\nn\\
+&\sum_{m=1}^{j}\sqrt{(j+m)(j-m+1)}\frac{g}{\sqrt{2}}W^-_\mu\bar\chi_{m-1}\gamma^\mu\chi_m\nn\\
+&{\rm h.c.}
\label{eq:lag2}
\end{align}
%\end{widetext}
As in the Dirac case, the electroweak quantum corrections spoil the degeneracy of the masses, making the neutral component as the lightest one.

For both Dirac and Majorana cases, when $j=1$, $\chi$ is an $SU(2)_L$ triplet, so $\bar{\chi}\cdot\phi E_L$ as an SM singlet can enter into the Lagrangian, where $\phi$ is the Higgs boson field, and $E_L$ stands for the left-handed lepton field. As a result, we have to introduce an extra $Z_2$ symmetry for the stability of DM candidate $\chi_0$, or assume $\chi$ does not carry lepton number as Ref.~\cite{Cirelli:2014dsa} shows. While when $j\geq2$, $\chi_0$ is automatically stable because decay modes consistent with renormalizability do not exist~\cite{Cirelli:2005uq}, therefore no other symmetry is needed.

The spin independent interaction rate of the DM with nuclei is suppressed in the MDM models, as it is generated only at the one-loop order~\cite{Hisano:2011cs,Hill:2011be,Hisano:2012wm,Hill:2013hoa,Hisano:2015rsa,Han:2018gej}. It is shown that for a triplet state, the spin-independent cross section with nucleons is only mildly sensitive on the DM mass and is around $10^{-47}\sim 10^{-49}~{\rm cm}^2$. It is dangerously close to the ``WIMP discovery limit'' imposed by the neutrino background, and the prospects for detection via the spin-independent direct detection are dim~\cite{Cirelli:2014dsa} and need constructing of larger and ultra-low noise detectors.

\section{Collider phenomenology}\label{sec:collider}

In this section we explore the discovery potential of MDM in proton-proton ($pp$) collisions. We first give an overview of our methods and then present our simulation results of various searching channels of DM at the 13~TeV Large Hadron Collider (LHC) and a 100~TeV $pp$ collider~\cite{Arkani-Hamed:2015vfh,CEPC-SPPCStudyGroup:2015csa,Golling:2016gvc,Contino:2016spe}. Unless otherwise specified, we consider an integrated luminosity ($\mathcal{L}$) of $3~{\rm ab}^{-1}$ at the LHC (named as HL-LHC) and an integrated luminosity of $30~{\rm ab}^{-1}$ at the 100~TeV collider. The connection of collider constraints and relic abundances will be built within the MDM models in the end.

\subsection{Overview}

A highly degeneration among the DM candidate and its weak partners are understood in this work. The dominant production channel of MDM in $pp$ collisions is through the so-called Drell-Yan process, in which a pair of dark sector particles are produced by mediating an EW gauge boson $W/Z/\gamma$. Unfortunately, due to the smallness of the mass splittings among dark particles in the MDM models, those processes result in both invisible particles (the DM candidates) and very soft ($\sim100$ MeV) visible particles (pions or leptons) in the final state. That is very hard to detect at high energy $pp$ colliders using the conventional detection technique. Recently a few novel strategies are proposed to search for those long-lived charged particles, e.g. the disappearing tracks~\cite{Low:2014cba,Cirelli:2014dsa,Ostdiek:2015aga,Han:2018wus}. Charged tracks of about 20~cm length could be detected after the Run-II upgrade of the ATLAS detector~\cite{Han:2018wus,Capeans:2010jnh}. We do not consider the detection of those long-lived charged particles in this work. Rather, we focus on the traditional searching strategies, i.e. the so-called mono-$X$ channel~\cite{Cao:2009uw,Cirelli:2014dsa,Goodman:2010ku,Fox:2011pm,An:2012ue,Haisch:2013ata,Low:2014cba,Han:2018wus}. In those channels the dark particles appear as large missing transverse energy ($\met$) in the detector, and they are produced in association with a detectable $X$ object, e.g. hard jet, $W/Z/\gamma$ or Higgs, etc.  Another powerful approach is the vector boson fusion (VBF) channel~\cite{Delannoy:2013ata,Brooke:2016vlw,Rauch:2016pai}, which has two energetic forward-backward jets and large $\met$ in the final state.
Obviously, the searches using the mono-$X$ signature and the long-lived charged particle will compliment one another.

From the viewpoint of collider phenomenology, the experimental sensitivity of the MDM models increases with the weak isospin of dark particles; the larger representation the dark particles exhibit, the larger coupling strengths and larger numbers of production channels. But from the viewpoint of cosmology, too many components of MDM would result in a large co-annihilation cross section in the early Universe, which will dramatically reduce the DM relic abundance. Therefore, we focus our attention to the weak isospin $j=1,~2,~3$ representations in the MDM models for both Dirac and Majorana fermions. We name them as ``D1" model for a Dirac DM with $j=1$, ``M1" model for a Majorana DM with $j=1$, and so on. 

To perform the collider phenomenology studies, we write the models described by Eq.~(\ref{eq:lag1}) and Eq.~(\ref{eq:lag2}) in UFO files~\cite{Degrande:2011ua} by use of {\tt FeynRules 2.0}~\cite{Alloul:2013bka}. For simplicity the mass splittings between dark particles are fixed to 200~MeV. The collider searching in the mono-$X$ signature is not affected by such a tiny mass split. Both signal and background events are generated using {\tt MadGraph5\_aMC@NLO}~\cite{Alwall:2014hca} at leading order with the {\tt NN23LO1}~\cite{Ball:2013hta} parton distribution functions (PDFs) at parton level, and then interfaced to {\tt Pythia 6.4}~\cite{Sjostrand:2006za} and {\tt Delphes 3}~\cite{deFavereau:2013fsa} for parton shower and fast detector simulation. We follow the the ATLAS~\cite{Aaboud:2017phn,Aaboud:2017dor} and CMS~\cite{Sirunyan:2017jix,Sirunyan:2017ewk,Khachatryan:2016mbu} collaborations to perform detailed simulations at the HL-LHC and 100~TeV collider. We find that the different strategies used by the ATLAS and CMS collaborations yield quite similar results in the end. For clarity and simplicity, we present the simulation result using the strategy of CMS collaboration throughout this paper.

\subsection{DM search in the Mono-jet channel }

\subsubsection{Collider simulation}

\begin{figure}[b]
\includegraphics[scale=0.8]{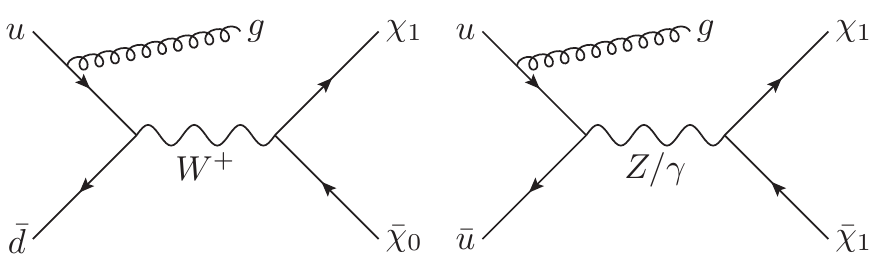}
\caption{Representative Feynman diagrams of the mono-jet channel in hadron collisions.}
\label{pic:monojet}
\end{figure}

We start with the mono-jet channel, which includes the production of a pair of MDM particles and an energetic light-flavor jet from the initial state radiation (ISR) of the parton. See Fig.~\ref{pic:monojet} for illustration, where $\chi_0$ denotes the DM candidate and $\chi_1$ represents other dark particles. We include all the production channels of mono-parton plus a pair of dark particles in this study. Such processes have been well studied in the literature, both theoretically~\cite{Cao:2009uw,Cirelli:2014dsa,Goodman:2010ku,Fox:2011pm,An:2012ue,Haisch:2013ata,Low:2014cba,Han:2018wus} and experimentally~\cite{Aad:2015zva,Aaboud:2017phn,Khachatryan:2014rra,Sirunyan:2017jix}. The event topology of the signal event consists of one hard jet plus large missing transverse momentum originating from the two dark particles in the final state. The dominant backgrounds are the $Z+{\rm jets}$ production with a subsequent decay of $Z\to\nu\bar\nu$ and the $W+{\rm jets}$ production with $W^\pm\to\ell^\pm\nu~(\ell=e,\mu,\tau)$. Other backgrounds,  e.g. $Z/\gamma^*(\to\ell^+\ell^-)$+jets, QCD multi-jets, $t\bar{t}$, single-top and di-boson $(W^+W^-,W^\pm Z,ZZ)$ processes, turn out to be negligible~\cite{Sirunyan:2017jix} and are ignored in our study.

In order to avoid the soft/collinear divergence from the QCD radiation when generating events at parton level, we demand all the light-flavor quarks or gluons from the ISR must exhibit a transverse momentum ($p_T$) larger than 10 GeV and appear in the central rapidity ($\eta$) region, say  $|\eta|<5$.  At the hadron level, to suppress the background events, we require at least one hard light-flavor jet in the central region of detector and also a significant missing transverse momentum ($\met$) at both the 13 TeV LHC and 100 TeV $pp$ colliders, i.e.
\beq
\label{eq:cut}
p_T^{L_j}>20~{\rm GeV},\quad |\eta^{L_j}| < 2.4,\quad \met>30~{\rm GeV},
\eeq
where $p_T^{L_j}$ and $\eta^{L_j}$ denotes the transverse momentum and rapidity of the leading jet ($L_j$).  In addition, a separation between the hard jet and missing energy in the azimuthal plane is required to satisfy $\Delta\phi(\vec{p}_T^{L_j},\vec{\slashed{p}}_T) > $ 0.5 radians. The CMS study shows that the QCD radiation effects can easily generate more than one jet in the signal and background events. In order to suppress those QCD radiation effect, the CMS collaboration further requires that  $\Delta\phi(\vec{p}_T^{j},\vec{\slashed{p}}_T) > $ 0.5 radians for those addition jets which exhibit a transverse momentum harder than 30~GeV.  

We name all the above cuts as ``selection cut" and display the numbers of signal of signal events ($n_s$) and background events ($n_b$) passing the cut in Table~\ref{tbl:cut_eff}. Note that we sum all the production channels of dark particles in the signal events and do not distinguish each individual channel. For illustration we choose $m_{\chi_0}=500~{\rm GeV}$ as the benchmark mass at the 13~TeV LHC and $m_{\chi_0}=1000~{\rm GeV}$ at the 100~TeV collider, both with an integrated luminosity ($\mathcal{L}$) of $100~{\rm fb}^{-1}$. For a given weak isospin of MDM, the production rate of Dirac DM exactly doubles the rate of Majorana DM after applying the generator cut, i.e. 
\beq
\frac{\sigma_{\rm D1}}{\sigma_{\rm M1}} = \frac{\sigma_{\rm D2}}{\sigma_{\rm M2}}=\frac{\sigma_{\rm D3}}{\sigma_{\rm M3}} \simeq 2.
\label{eq:ratio1}
\eeq
It is owing to the fact that the kinematics of the signal events is controlled solely by the DM mass rather than their fermionic feature, either of Dirac or Majorana type. 
Also, the ratios of production rates in the MDM models are 
\bea
&&\frac{\sigma_{\rm D2}}{\sigma_{\rm D1}}\simeq \frac{\sigma_{\rm M2}}{\sigma_{\rm M1}}\sim 5,\nn\\
&&\frac{\sigma_{\rm D3}}{\sigma_{\rm D1}}\simeq \frac{\sigma_{\rm M3}}{\sigma_{\rm M1}} \sim 14,
\label{eq:ratio2}
\eea
at both the 13~TeV LHC and the 100~TeV collider after the generator and veto cuts. See the second row and the seventh row in Table~\ref{tbl:cut_eff}. The quintet and septet MDMs, which have larger strength of weak couplings and more production channels than the triplet MDM, exhibit a larger production rate and are easy to probe at hadron colliders.

The SM background processes often consist of charged leptons and heavy flavor ($b$) jets. Following the CMS collaboration~\cite{Sirunyan:2017jix} we apply a set of veto cuts to remove those charged leptons, photons and $b$-jets in the reducible backgrounds. The veto cuts are listed as follows: 
\bea
e^\pm &:\qquad & p_T^e >10~{\rm GeV}, \qquad |\eta^e| < 2.5,\nn\\
\mu^\pm &:\qquad & p_T^{\mu} >10~{\rm GeV}, \qquad |\eta^{\mu}| <2.4,\nn\\
\tau^\pm &:\qquad & p_T^\tau >18~{\rm GeV},\qquad |\eta^\tau| <2.3,\nn\\
\text{photon}~(\gamma) &:\qquad & p_T^\gamma >15~{\rm GeV},\qquad|\eta^\gamma| <2.5,\nn\\
b\text{-jet} &:\qquad & p_T^b >20~{\rm GeV},\qquad |\eta^b| <2.4~.
\eea
To expedite the MC simulation, we demand the $Z$ boson in the $Z+{\rm jets}$ background decays into a pair of neutrinos rather than charged leptons when generating the background events in {\tt MadGraph5\_aMC@NLO}. Hence, the veto cuts slightly reduce the rate of $Z+{\rm jets}$ background. On the other hand, the $W^\pm$ boson in the $W^\pm +{\rm jets}$ background is required to decay leptonically, i.e. $W^\pm \to \ell^\pm \nu~(\ell=e,\mu,\tau)$, which yields a charged lepton and a neutrino in the final state. At the LHC the $W^\pm$ boson is boosted such that the charged lepton from its decay often populates in the forward region to escape the veto cut. As a result, after the veto cuts the rate of $W^\pm + {\rm jets}$ background is reduced by a factor of about 2. See the third row and twelfth row of Table~\ref{tbl:cut_eff}.  Our simulation results of the $Z+{\rm jets}$ and the $W^\pm+{\rm jets}$ backgrounds are in a good agreement with those given by the CMS collaboration~\cite{Sirunyan:2017jix} at the $\sqrt{s}=13$ TeV LHC with an integrated luminosity of 35.9 fb$^{-1}$.

\begin{table*}
\caption{The number of signal and background events at the 13~TeV LHC with $m_{\chi_0}=500~{\rm GeV}$ (top) and 100~TeV collider with $m_{\chi_0}=1000~{\rm GeV}$ (bottom) with an integrated luminosity of $100~{\rm fb}^{-1}$. The statistical significance ($\mathcal{S}$) of each individual scenario is also shown. Three cut scenarios are examined to optimize the significance of signal events; see Eqs.~(\ref{eq:cut13}) and ~(\ref{eq:cut100}). }
\label{tbl:cut_eff}
\begin{ruledtabular}
\begin{tabular}{l|cccccc|ccc}
\multicolumn{1}{l|}{13~TeV LHC}     &   \multicolumn{6}{c|}{Signal}   &         \multicolumn{3}{c}{Background}       \\
$M_\chi=500$ GeV &  D1   &  D2   &  D3      &  M1      &  M2      &  M3      &  $Z$+jets    &  $W^\pm$+jets     &  All          \\
\colrule
Selection cut              &  7593    &  37904   &  105430  &  3813    &  18895   &  52878   &  376860500   &  1872116000   &  2248976500   \\
Veto cut           &  7392    &  36256   &  100924  &  3644    &  18413   &  51482   &  359756000   &  1008355000   &  1368111000   \\ \hline
Cut-1              &  443     &  2150    &  5767    &  207     &  1108    &  3187    &  423657      &  196429       &  620086       \\
$\mathcal{S}$     &  0.563   & 2.727   &  7.301  & 0.263   &  1.406  & 4.040   &              &               &               \\
\hline
Cut-2              &  78      &  403     &  1082    &  36      &  190     &  555     &  12497       &  3354         &  15851        \\
$\mathcal{S}$      &  0.621  &  3.175   &  8.408   &  0.287  &  1.502   &   4.357   &              &               &               \\
\hline
Cut-3              &  28      &  148     &  407     &  12      &  71      &  199     &  2814        &  709          &  3523         \\
$\mathcal{S}$     &  0.462   &  2.467  & 6.613  & 0.203   &  1.181  &  3.295   &              &               &               \\
\hline
\colrule
\hline
\colrule
\multicolumn{1}{l|}{100~TeV}     &   \multicolumn{6}{c|}{Signal}   &         \multicolumn{3}{c}{Background}       \\
$M_\chi=1000$ GeV &  D1   &  D2   &  D3      &  M1      &  M2      &  M3      &  $Z$+jets    &  $W^\pm$+jets     &  All          \\
\colrule
Selection cut              &  29841   &  149110   &  417010 &  15073   &  74904   &  209110  &  3911173333  &  18976213142  &  22887386476  \\
Veto cut           &  27713   &  138466   &  388068 &  14055   &  69655   &  194107  &  3690570000  &  12284410000  &  15974980000  \\
\hline
Cut-I              &  2577    &  13041    &  36186  &  1310    &  6513    &  18451   &  2262530     &  1319165      &  3581695      \\
$\mathcal{S}$     & 1.361   &  6.882    &  19.056 &  0.692   & 3.439   &  9.733   &              &               &               \\
\hline
Cut-II             &  2065    &  10650    &  29037  &  1056    &  5267    &  14893   &  849077      &  286632       &  1135709      \\
$\mathcal{S}$    & 1.936   & 9.962   & 27.018 &  0.990   &  4.935   & 13.914  &              &               &               \\
\hline
Cut-III            &  678     &  3580     &  9325   &  337     &  1687    &  4828    &  41868       &  2817         &  44684        \\
$\mathcal{S}$  & 3.193   &  16.503   & 41.373 & 1.590    &  7.883   &   22.063 &              &               &               \\
\end{tabular}
\end{ruledtabular}
\end{table*}

After the selection and veto cuts, the background is much larger than the signal; see the third row in Table~\ref{tbl:cut_eff}.  In order to suppress the huge background, we impose strong cuts on the leading jet and $\met$, named as ``optimal cuts".  For the analysis of 13~TeV LHC we choose the cut thresholds close to those used by the CMS collaboration~\cite{Sirunyan:2017jix} and further vary the $\met$ cut to optimize the signal. The following three scenarios of optimal cuts are used in our analysis
\bea
\text{cut-1}&:\quad & p_T^{L_j}\geqslant100~{\rm GeV},\quad \met\geqslant250~{\rm GeV},\nn\\
\text{cut-2}&:\quad& p_T^{L_j}\geqslant100~{\rm GeV},\quad \met\geqslant500~{\rm GeV},\nn\\
\text{cut-3}&:\quad & p_T^{L_j}\geqslant100~{\rm GeV},\quad \met\geqslant700~{\rm GeV}.
\label{eq:cut13}
\eea
While at the 100~TeV collider we choose a much harder cut on the leading jet to suppress the enormous QCD backgrounds, e.g. 
\bea
\text{cut-I}&:\quad & p_T^{L_j}\geqslant400~{\rm GeV},\quad \met\geqslant200~{\rm GeV},\nn\\
\text{cut-II}&:\quad& p_T^{L_j}\geqslant400~{\rm GeV},\quad \met\geqslant500~{\rm GeV},\nn\\
\text{cut-III}&:\quad & p_T^{L_j}\geqslant400~{\rm GeV},\quad \met\geqslant1000~{\rm GeV}.
\label{eq:cut100}
\eea
We consider all the three cuts on $\met$ in our study and  choose the best one for discovery or exclusion for each DM mass in a specific MDM model. Table~\ref{tbl:cut_eff} displays the numbers of signal and background events after imposing the optimal cuts and the corresponding statistical significance $\mathcal{S}(\equiv n_s/\sqrt{n_b})$ at the two colliders with an integrated luminosity of $100~{\rm fb}^{-1}$. Here, we assume the signal and background events obeying the Gaussian statistics such that results of other luminosities can be easily obtained by rescaling those numbers shown in the table. Note that we use a more proper Possion statistics when estimating the discovery and exclusion potential of the HL-LHC and 100~TeV collider. We notice that the optimal cuts reduce the SM background significantly; for example, the background events are suppressed by a factor of $4\times 10^{-4}$ ($10^{-5}$, $2\times 10^{-6}$) after the cut-1 (cut-2, cut-3), respectively, at the 13~TeV LHC. On the other hand,  the signal events are reduced by a factor of 0.06 (0.01, 0.004), respectively. That increases the signal-to-background ratio greatly. The hardest cut on $\met$ in the optimal cuts yields the best of discovery significance but inevitably leads to fewer numbers of events. 

\begin{figure}[b]
\includegraphics[scale=0.23]{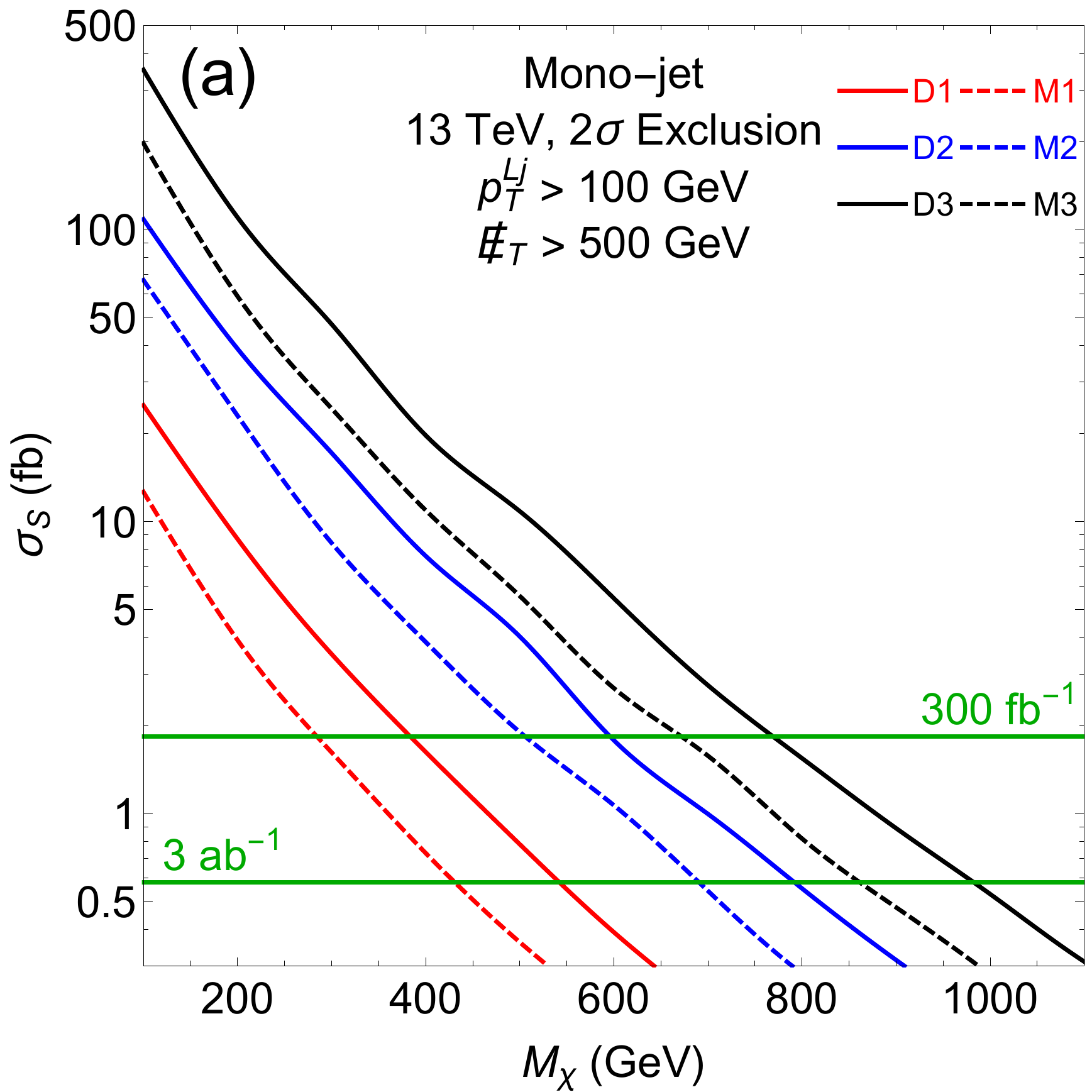}
\includegraphics[scale=0.23]{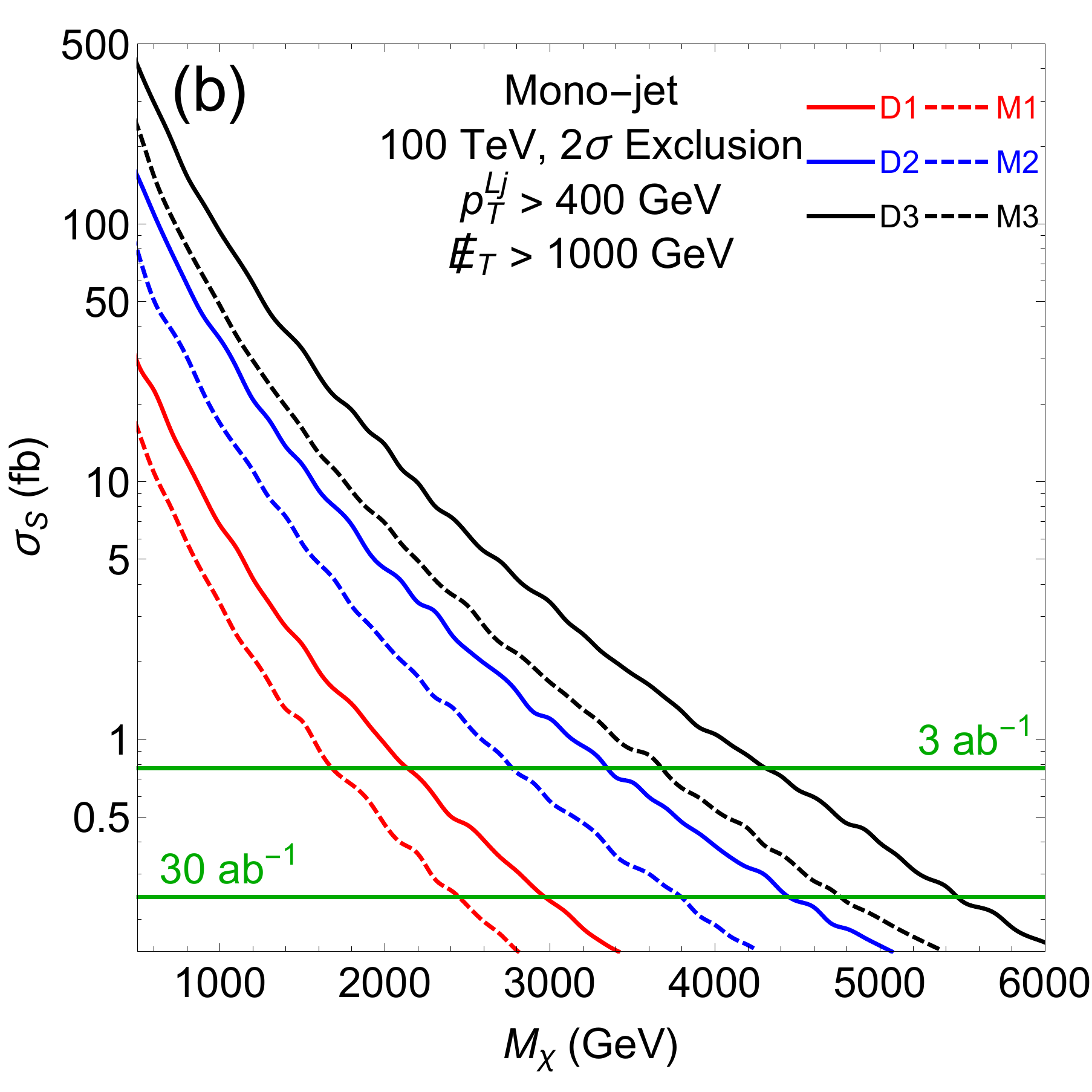}
\caption{The cross section of the mono-jet signal process as a function of $M_\chi$ after imposing the cut-2 at the 13~TeV LHC (a) and after the cut-III at the 100~TeV collider (b). The solid curves represent the signal cross section of Dirac type DM while the dashed curves the Majorana type DM. The red (blue, black) curve denotes the weak triplet (quintet, septet) DM, respectively. The horizontal green curves represent the $2\sigma$ exclusion limits with the chosen benchmark integrated luminosities.}
\label{pic:monojet_sigmas}
\end{figure}

Figure~\ref{pic:monojet_sigmas} plots the production cross section of the signal event as a function of the DM mass $m_{\chi_0}$ after the cut-2 at the 13~TeV (a) and the cut-III at the 100~TeV collider (b). The ratios of production rates in the MDM models remain almost the same as those in Eqs.~(\ref{eq:ratio1}) and~(\ref{eq:ratio2}) for all the three hard cuts, e.g. 
\bea
&& \frac{\sigma_{\rm D1}}{\sigma_{\rm M1}} = \frac{\sigma_{\rm D2}}{\sigma_{\rm M2}}=\frac{\sigma_{\rm D3}}{\sigma_{\rm M3}} \simeq 2,\nn\\
&&\frac{\sigma_{\rm D2}}{\sigma_{\rm D1}}\simeq \frac{\sigma_{\rm M2}}{\sigma_{\rm M1}}\sim 5\sim 6,\nn\\
&&\frac{\sigma_{\rm D3}}{\sigma_{\rm D1}}\simeq \frac{\sigma_{\rm M3}}{\sigma_{\rm M1}} \sim 13\sim 15.
\eea
Equipped with the optimal cuts shown above, we estimate the region of $m_{\chi_0}$ to claim a 5 standard deviations ($\sigma$) statistical significance using 
\beq
\sqrt{-2\left[(n_b + n_s) \log\frac{n_b}{n_s+n_b}+n_s\right]}=5,
\label{eq:5sigma}
\eeq
for a given integrated luminosity $\mathcal{L}$. In the case that no evidence of the DM candidate is observed, one can set a $2\sigma$ exclusion limit on $m_{\chi_0}$ from
\beq
\sqrt{-2\left[n_b\log \frac{n_s + n_b}{n_b}-n_s\right]} = 2~.
\label{eq:2sigma}
\eeq
For each individual optimal cut we first obtain the $2\sigma$ exclusion limit on $n_s$ from the number of background event $n_b$ and then derive the corresponding bound on the cross sections of the signal from $n_s/\mathcal{L}$. For example, the horizontal green curves in Fig.~\ref{pic:monojet_sigmas} represent the $2\sigma$ exclusion bounds on $\sigma_s$ for the benchmark $\mathcal{L}$ at the two colliders. The lower limits of $m_{\chi_0}$ follow from the intersection points between the $\sigma_s$ and exclusion bound curves.

\begin{table}[b]
\caption{Exclusion limits of the DM mass (in unit of GeV) from the mono-jet channel at the 13TeV LHC with $\mathcal{L}=3~{\rm ab}^{-1}$ and at the 100 TeV collider $\mathcal{L}=30~{\rm ab}^{-1}$.}
\label{tab:monojet_cms2}
\begin{tabular}{lcccccc}
\hline\hline
13 TeV ($3~{\rm ab}^{-1}$) &D1&D2&D3&M1&M2&M3\\
\colrule
Cut-1             & 531  &  767   &  935  &  434    &  668   &  832   \\
Cut-2            & 544 &  794  &  983 &  432   &  691&  864  \\
Cut-3           & 522 &  781 &  989   &  391 &  682 &  853  \\
\hline
The best  & 544 &  794  &  989   &  434   &  691  &  864   \\
\hline\hline\\
~\\
\hline
\hline
100 TeV ($30~{\rm ab}^{-1}$)&D1&D2&D3&M1&M2&M3\\
\colrule
Cut-I     & 2161 &  3305  &  4203  &  1768   &  2778  &  3597   \\
Cut-II   & 2419  &  3676 &  4630 &  1993  &  3070 &  3956 \\
Cut-III   & 2970 &  4465 &  5479  &  2449  &  3797  &  4772  \\
\hline
The best & 2970&  4465   &  5479  &  2449  &  3797  &  4772   \\
\hline
\hline
\end{tabular}
\end{table}

Table~\ref{tab:monojet_cms2} shows the exclusion limits on $m_{\chi_0}$ at the 95\% confidence level for various optimal cuts at the HL-LHC  with $\mathcal{L}=3~{\rm ab}^{-1}$ (top panel) and at the 100~TeV collider with $\mathcal{L}=30~{\rm ab}^{-1}$ (bottom panel). The last row of each panel displays the best exclusion limit for each MDM model.  It shows that, at the 13~TeV LHC, the cut-1 scenario works the best for the M1 model, the cut-2 yields the best exclusion limits for the D1, D2, M2 and M3 models, and the cut-3 works the best for the D3 model. On the other hand, the cut-3 (the hardest cut) is the best choice for all the MDM models at the 100~TeV collider. Note that the exclusion limit of $m_{\chi_0}$ increases with the quantum number of dark particles under the $SU(2)_L$ group, owing to the large production rate. For example, the limit of the D3 model is nearly twice bigger than that of the D1 model. Both the Dirac and Majorana type DM candidates, when they share equal weak quantum numbers, yield comparable bounds.  

Note that the current 13~TeV LHC searches impose a lower bound of 460~GeV for the DM candidate in the  D1 model at the 95\% C.L.~\cite{Aaboud:2017mpt} based on the disappearing-track signature and an integrated luminosity of $36.1~{\rm fb}^{-1}$. The bound is slightly weaker than our result $m_{\chi_0}\geqslant544~{\rm GeV}$ (assuming an integrated luminosity of $3~{\rm ab}^{-1}$). We expect that collecting more data sample the disappearing-track signature would give a much stronger bound than the mono-jet channel.  

Our results are compatible with Refs.~\cite{Cirelli:2014dsa,Han:2018wus}, which explore the collider phenomenology of wino-like DM that corresponds to our M1 case. For example, Ref.~\cite{Han:2018wus} presents the exclusion limits of 280 (900) GeV for a wino-like DM candidate at the 14 TeV HL-LHC with $\mathcal{L}=3~{\rm ab}^{-1}$ and 2 (6.5) TeV at the 100 TeV collider with $\mathcal{L}= 30 ~{\rm ab}^{-1}$, using the mono-jet (disappearing charged tracks) search, respectively. Here in our study, we exclude the triplet Majorana MDM up to $\sim 400$ GeV at the 13 TeV 3 ab$^{-1}$ LHC, and $\sim 2$ TeV at the 100 TeV 30 ab$^{-1}$ $pp$ collider using the mono-jet search. While the disappearing tracks approach performs better, it is complicated for the estimation of the SM backgrounds at future colliders~\cite{Cirelli:2014dsa,Han:2018wus}, therefore, the traditional method is easy to implement and also provide a crosscheck of the novel method.

\subsubsection{Limiting relic abundance from colliders}

Now we connect the collider searches with the relic abundance in the early Universe in the MDM model. 

Once the weak quantum number of DM particles is given, there is only one parameter in the MDM model, i.e. the DM mass $m_{\chi_0}$. The relic abundance is inversely proportional to the DM annihilation or co-annihilation cross sections which, from dimension analysis, can be quantitively written as 
\beq
\sigma(\chi\bar{\chi} \to XY)_{\rm ann} \sim  \frac{g^4}{M_{\chi}^2}. 
\eeq
where $X$ and $Y$ denotes the SM particles whose mass effects are ignored here.  The relic abundance can be approximately given by 
\beq
\Omega h^2 \Big|_{M_\chi}\sim \frac{0.1~\text{pb}}{\sum_i\left<\sigma v\right>_{\rm ann}^i} \sim \frac{M_\chi^2}{g^4},
\eeq
i.e. the relic abundance is proportional to the DM mass square.  After knowing the exclusion limit of $m_{\chi_0}$ given by the collider searches, say $m_{\chi_0}\geqslant\mathcal{M}_\chi^{\rm min}$, we calculate the lower bound on the relic abundance of DM as 
\beq
\Omega h^2_{\chi}\equiv \Omega h^2\Big|_{\mathcal{M}_\chi^{\rm min}}\sim \frac{\left(\mathcal{M}_\chi^{\rm min}\right)2}{g^4}. 
\eeq
Define $m_{\chi}^{\rm C}$ as the critical DM mass that generates the observed relic abundance ($\Omega h^2_{\rm DM}\simeq 0.12$), i.e. 
\beq
\Omega h^2_{\rm DM}\equiv \Omega h^2\Big|_{m_\chi^{\rm C}} \simeq 0.12\sim \frac{(m_\chi^{\rm C})^2}{g^4},
\eeq
where $m_\chi^{\rm C}$'s in the MDM models are given by
\begin{align}
&\text{D1:~~} m_\chi^{\rm C}= 1.6~\text{TeV};   &&\text{M1:~~}  m_\chi^{\rm C}=  2.3~\text{TeV};\nn\\
&\text{D2:~~} m_\chi^{\rm C}= 3.0~\text{TeV};   &&\text{M2:~~}  m_\chi^{\rm C}= 4.3~\text{TeV};\nn\\
&\text{D3:~~} m_\chi^{\rm C}= 4.6~\text{TeV};   &&\text{M3:~~}  m_\chi^{\rm C}= 6.6~\text{TeV}.
\end{align}
We further define the fraction of the DM relic abundance predicted by the exclusion limit of the mono-jet search inside the observed relic abundance as  
\beq
\mathcal{F}\equiv \frac{\Omega_\chi}{\Omega_{\rm DM}} \simeq \left(\frac{\mathcal{M}_\chi}{m_\chi^{\rm C}}\right)^2,
\eeq
i.e. the quota of DM inside the observed relic abundance $\Omega h^2_{\rm DM}~(\simeq 0.1)$. 
Note that the DM mass ($m_\chi^c$) yielding the observed relic abundance is fixed in a specific MDM model, therefore, the fraction defined above could tell us whether or not the MDM is adequate to explain the relic abundance. For example, if $\mathcal{F}<1$, then the MDM cannot explain the relic abundance and extra candidate of DM is needed to compensate the deficit; if $\mathcal{F}> 1$, then the MDM overproduces the relic abundance and additional annihilation mechanics is needed to reduce the amount of DM. 

\begin{figure}
\includegraphics[scale=0.35]{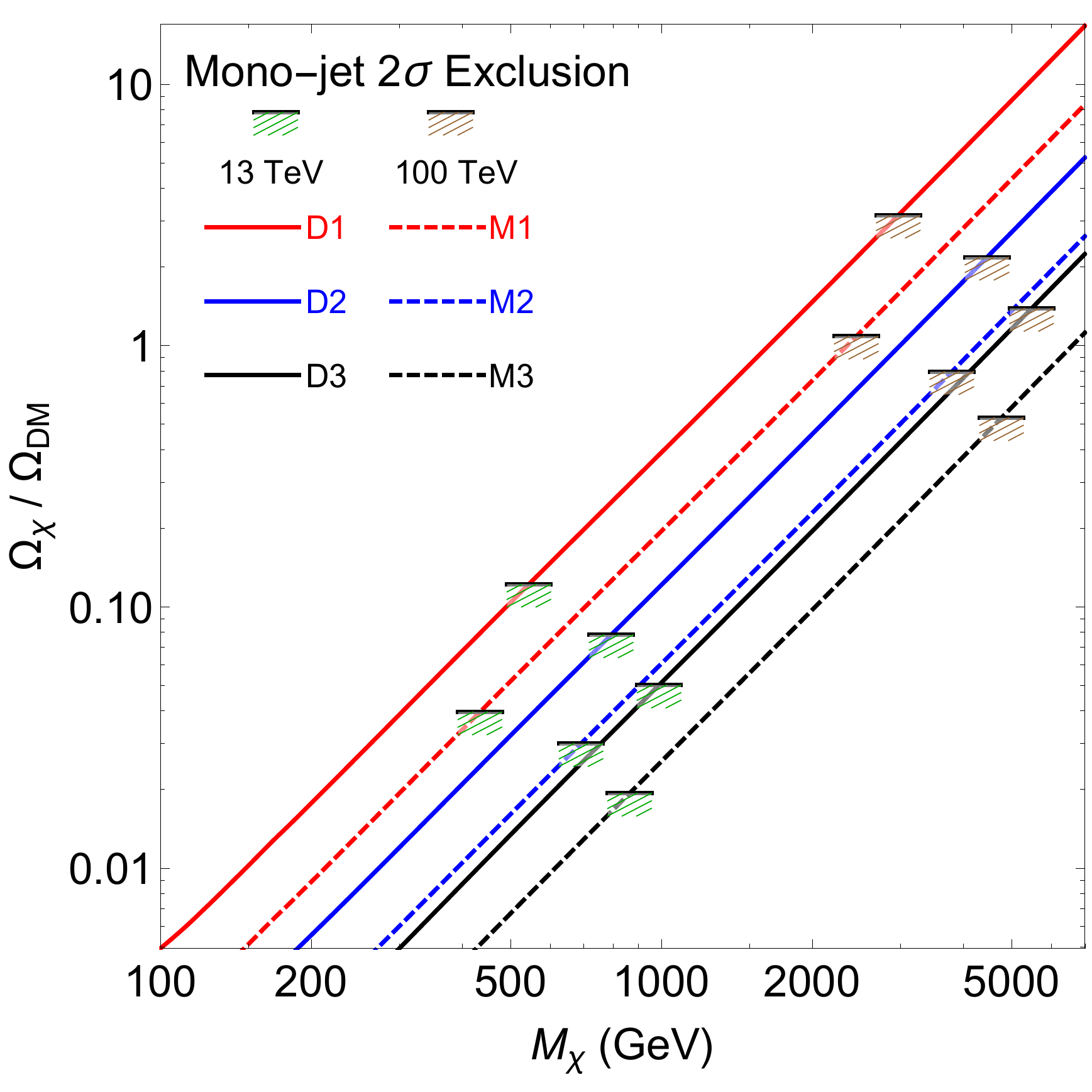}
\caption{The fraction $\mathcal{F}=\Omega_\chi/\Omega_{\rm DM}$ as a function of $m_{\chi_0}$ in the MDM models where the horizontal minibars label the lower limits of $m_{\chi_0}$ given by our simulation of the mono-jet channel at the HL-LHC ($\mathcal{L}=3~{\rm ab}^{-1}$) and the 100~TeV collider ($\mathcal{L}=30~{\rm ab}^{-1}$).
}
\label{pic:monojet_results}
\end{figure}

We calculate the relic abundance for different DM masses in the MDM models using {\tt micrOMEGAs}~\cite{Belanger:2018mqt} and then plot the fraction $\mathcal{F}$ curves as a function of $m_{\chi_0}$ in Fig.~\ref{pic:monojet_results}: the solid curves denote the Dirac type DM while the dashed curves the Majorana DM. The red (D1 and M1), blue (D2 and M2),  and black (D3 and M3) curve denotes the triplet, quintet and septet DM, respectively. The horizontal minbars represent the collider constraints and the intersections of the minbars and the relic abundance curves mark the lower bounds of $m_{\chi_0}$ obtained from our simulation.  The region of the fraction $\mathcal{F}$ above the minibar is allowed.  From the $2\sigma$ exclusion bounds on $m_{\chi_0}$ shown in Table~\ref{tab:monojet_cms2} we obtain the lower limits of the fraction $\mathcal{F}$
%as shown in Fig.~\ref{pic:monojet_results}(b) in which the shaded region represents the fraction $\mathcal{F}$ excluded by the DM search in the mono-jet channel at the HL-LHC (left panel) and 100~TeV collider (right panel). If a null result is found in the mono-jet search, then we can exclude the fraction $\mathcal{F}$ 
at the HL-LHC as follows: 
\begin{align}
&\text{D1:~~} \mathcal{F}\geqslant12.2\%, &&\text{M1:~~}  \mathcal{F}\geqslant4.0\%,\nn\\
&\text{D2:~~} \mathcal{F}\geqslant7.8\%,   &&\text{M2:~~}  \mathcal{F}\geqslant3.0\%,\nn\\
&\text{D3:~~} \mathcal{F}\geqslant5.0\%,   &&\text{M3:~~}  \mathcal{F}\geqslant1.9\%,
\end{align}
while at the 100~TeV collider ($\mathcal{L}=30~{\rm ab}^{-1}$)
\begin{align}
&\text{D1:~~} \mathcal{F}\geqslant314.9\%,  &&\text{M1:~~}  \mathcal{F}\geqslant108.4\%,\nn\\
&\text{D2:~~} \mathcal{F}\geqslant217.5\%,   &&\text{M2:~~}  \mathcal{F}\geqslant79.3\%,\nn\\
&\text{D3:~~} \mathcal{F}\geqslant138.5\%,   &&\text{M3:~~}  \mathcal{F}\geqslant52.9\%.
\end{align}
Obviously, there is a significant promotion for the relic abundance constraints between HL-LHC and the 100 TeV collider. The Dirac type DMs have better constraints than the Majorana DMs, due to the larger number of degrees of freedom in the spinor space. At the future 100 TeV collider, the MDM model that yields $\mathcal{F}>100\%$, such as D1, D2, D3 and M1, is excluded at the 95\% C.L. if we assume $\chi$ is the only source of DM.

\begin{table*}
\caption{The coefficients of DM annihilation and co-annihilation channels in the non-relativistic expansion for Dirac type DMs. The coefficients of Majorana type DMs are exactly half of those of Dirac DMs. The branching ratio of annihilation channels are also shown.  }
\label{tbl:dm_ann}
\renewcommand\arraystretch{1.8}
\setlength{\tabcolsep}{2mm}{
\begin{tabular}{l|c|c|c|c|c|c|c}
\hline 
\multicolumn{2}{c|}{} & \multicolumn{2}{c|}{D1} & \multicolumn{2}{c|}{D2} & \multicolumn{2}{c}{D3}\tabularnewline
\cline{3-8} 
\multicolumn{2}{c|}{} & Coefficents & Br & Coefficents & Br & Coefficients & Br\tabularnewline
\hline 
\multirow{2}{*}{$\chi\bar{\chi} \to q\bar{q}$} & $a$ & $\dfrac{27 g^4}{16 \pi  M_{\chi }^2}$  & \multirow{2}{*}{47.9\%} &  $\dfrac{135 g^4}{16 \pi  M_{\chi }^2}$  & \multirow{2}{*}{25.4\%} &  $\dfrac{189 g^4}{8 \pi  M_{\chi }^2}$  & \multirow{2}{*}{15.0\%}\tabularnewline
& $b$ & $-\dfrac{45 g^4}{128 \pi  M_{\chi }^2}$  &  & $-\dfrac{225 g^4}{128 \pi  M_{\chi }^2}$  &  & $-\dfrac{315 g^4}{64 \pi  M_{\chi }^2}$  & \tabularnewline
\hline 
\multirow{2}{*}{$\chi\bar{\chi} \to \ell\bar{\ell}$} & $a$ & $\dfrac{9 g^4}{16 \pi  M_{\chi }^2}$  & \multirow{2}{*}{16.0\%} & $\dfrac{45 g^4}{16 \pi  M_{\chi }^2}$ & \multirow{2}{*}{8.5\%} &  $\dfrac{63 g^4}{8 \pi  M_{\chi }^2}$ & \multirow{2}{*}{5.0\%}\tabularnewline
& $b$ & $-\dfrac{15 g^4}{128 \pi  M_{\chi }^2}$  &  & $-\dfrac{75 g^4}{128 \pi  M_{\chi }^2}$ &  & $-\dfrac{105 g^4}{64 \pi  M_{\chi }^2}$  & \tabularnewline
\hline 
\multirow{2}{*}{$\chi\bar{\chi} \to WW,W\gamma,WH$} & $a$ & $\dfrac{16 e^2 g^2+43 g^4}{64 \pi  M_{\chi }^2}$ & \multirow{2}{*}{21.8\%} & $\dfrac{336 e^2 g^2+727 g^4}{64 \pi  M_{\chi }^2}$  & \multirow{2}{*}{40.1\%} & $\dfrac{7 \left(144 e^2 g^2+299 g^4\right)}{32 \pi  M_{\chi }^2}$ & \multirow{2}{*}{49.0\%}\tabularnewline
& $b$ & $\dfrac{176 e^2 g^2+361 g^4}{1536 \pi  M_{\chi }^2}$  &  &$\dfrac{3184 e^2 g^2+6413 g^4}{1536 \pi  M_{\chi }^2}$  &  & $\dfrac{7 \left(1328 e^2 g^2+2665 g^4\right)}{768 \pi  M_{\chi }^2}$ & \tabularnewline
\hline 
\multirow{2}{*}{$\chi\bar{\chi} \to ZZ,Z\gamma,ZH$} & $a$ & $\dfrac{17 g^4-16 e^4}{64 \pi  M_{\chi }^2}$  & \multirow{2}{*}{7.4\%} &  $\dfrac{277 g^4-272 e^4}{64 \pi  M_{\chi }^2}$  & \multirow{2}{*}{12.4\%} & $\dfrac{7 \left(113 g^4-112 e^4\right)}{32 \pi  M_{\chi }^2}$   & \multirow{2}{*}{14.5\%}\tabularnewline
& $b$ &  $\dfrac{139 g^4-144 e^4}{1536 \pi  M_{\chi }^2}$ &  & $\dfrac{2423 g^4-2448 e^4}{1536 \pi  M_{\chi }^2}$  &  & $\dfrac{7 \left(1003 g^4-1008 e^4\right)}{768 \pi  M_{\chi }^2}$  & \tabularnewline
\hline 
\multirow{2}{*}{$\chi\bar{\chi} \to WZ$ } & a &$\dfrac{9 g^4-8 e^2 g^2}{32 \pi  M_{\chi }^2}$  & \multirow{2}{*}{6.6\%} & $\dfrac{173 g^4-168 e^2 g^2}{32 \pi  M_{\chi }^2}$  & \multirow{2}{*}{13.0\%} & $\dfrac{7 \left(73 g^4-72 e^2 g^2\right)}{16 \pi  M_{\chi }^2}$ & \multirow{2}{*}{15.8\%}\tabularnewline
& $b$ & $\dfrac{83 g^4-88 e^2 g^2}{768 \pi  M_{\chi }^2}$ &  & $\dfrac{1567 g^4-1592 e^2 g^2}{768 \pi  M_{\chi }^2}$  &  & $\dfrac{7 \left(659 g^4-664 e^2 g^2\right)}{384 \pi  M_{\chi }^2}$   & \tabularnewline
\hline 
\multirow{2}{*}{$\chi\bar{\chi} \to \gamma\gamma$ } & $a$ & $\dfrac{e^4}{4 \pi  M_{\chi }^2}$ & \multirow{2}{*}{0.4\%} & $\dfrac{17 e^4}{4 \pi  M_{\chi }^2}$ & \multirow{2}{*}{0.7\%} & $\dfrac{49 e^4}{2 \pi  M_{\chi }^2}$  & \multirow{2}{*}{0.8\%}\tabularnewline
& $b$ & $\dfrac{3 e^4}{32 \pi  M_{\chi }^2}$ &  & $\dfrac{51 e^4}{32 \pi  M_{\chi }^2}$  &  & $\dfrac{147 e^4}{16 \pi  M_{\chi }^2}$  & \tabularnewline
\hline \hline
\multirow{2}{*}{All channels} & $a$ & $\dfrac{111 g^4}{32 \pi  M_{\chi }^2}$  & \multirow{2}{*}{100\%} & $\dfrac{1035 g^4}{32 \pi  M_{\chi }^2}$ & \multirow{2}{*}{100\%} & $\dfrac{2457 g^4}{16 \pi  M_{\chi }^2}$  & \multirow{2}{*}{100\%}\tabularnewline
& $b$ & $-\dfrac{9 g^4}{256 \pi  M_{\chi }^2}$ &  &$\dfrac{1395 g^4}{256 \pi  M_{\chi }^2}$  &  &$\dfrac{4977 g^4}{128 \pi  M_{\chi }^2}$   & \tabularnewline
\hline 
\end{tabular}
}
\end{table*}

The relic abundance curves shown in Fig.~\ref{pic:monojet_results} can be understood as follows. Given the fact that DM particles are non-relativistic when they decouple from the thermal bath, the DM annihilation cross section $\ave{\sigma v}$ is well approximated by a non-relativistic expansion (obtained by replacing the square of the energy in the center of mass frame by $s=4M_\chi^2+M_\chi^2v^2$):~\cite{Kolb:1990vq,Bertone:2004pz,Bertone:2010zza}
\beq\label{annihilation}
\ave{\sigma v}=a+b\ave{v^2}+\mathcal{O}(\ave{v^4})\approx a+6b\times \frac{T_F}{M_{\rm DM}},
\eeq
where $T_F$ being the freeze-out temperature. As the DM of interest to us is very heavy, we can treat the SM particles as massless in the calculation of DM annihilation and co-annihilation processes. That yields quite simple expressions of the $a$ and $b$ terms for each annihilation process as shown in Table~\ref{tbl:dm_ann}. For simplicity we do not distinguish the annihilation and co-annihilation channels and sum them up as a single subcategory if they contribute to the same final state. We also show the branching ratio of each subcategories in the total annihilation channels. Table~\ref{tbl:dm_ann} only shows the results of Dirac type of DMs. The coefficients of Majorana type DMs are exactly half of those of Dirac DMs, and both the Dirac and Majorana type DMs, if they carry the same weak isospin, share the same branching ratios. Note that the quark mode dominates for the triplet DMs (D1 or M1) as benefiting from the color numbers of the quarks in the final state. On the other hand, the weak boson modes dominate for the quintet DMs (D2 or M2) and the septet (D3 or M3) DMs as the numbers of annihilation channels increase with the weak isospin quantum number so as to exceed the quark modes.    

After summing all the annihilation and co-annihilation channels, we obtain the coefficients $a$ and $b$ of all the channels as follows: 
\begin{align}
& {\rm D1:~} \qquad a=\frac{111 g^4}{32 \pi  M_{\chi }^2}, \qquad b=-\frac{9 g^4}{256 \pi  M_{\chi }^2}; \nn\\
&{\rm D2:~} \qquad a=\frac{1035 g^4}{32 \pi  M_{\chi }^2}, \qquad b=~\frac{1395 g^4}{256 \pi  M_{\chi }^2};\nn\\
&{\rm D3:~} \qquad a=\frac{2457 g^4}{16 \pi  M_{\chi }^2}, \qquad b=~\frac{4977 g^4}{128 \pi  M_{\chi }^2};\nn\\
&{\rm M1:~} \qquad a=\frac{111 g^4}{64 \pi  M_{\chi }^2},  \qquad b=-\frac{9 g^4}{512 \pi  M_{\chi }^2};\nn\\
&{\rm M2:~} \qquad a=\frac{1035 g^4}{64 \pi  M_{\chi }^2}, \qquad b=~\frac{1395 g^4}{512 \pi  M_{\chi }^2};\nn\\
&{\rm M3:~} \qquad a=\frac{2457 g^4}{32 \pi  M_{\chi }^2}, \qquad b=~\frac{4977 g^4}{256 \pi  M_{\chi }^2}.
\end{align}
It is obvious that, in each MDM model, the $b$-term is much smaller than the $a$-term and its contribution is further suppressed by the factor $T_F/M_{\chi}\sim 1/20$. Therefore, we can safely ignore the $b$-term in the discussion of those relic abundance curves.

We can see that, for each model, $a$ and $b$, as well as $\ave{\sigma v}$, are proportional to $M_\chi^{-2}$, so the relic abundance $\Omega_\chi h^2$ is proportional to $M_\chi^2$, which explains the linear behavior of $\mathcal{F}$-$M_\chi$ relationship in Fig.~\ref{pic:monojet_results}. Note that the $y$-axis is in lograthemic scale.  Furthermore, for both the Dirac and Majorana DMs, when $M_\chi$ is fixed, the bigger $j$ is, the larger $\ave{\sigma v}$ we get, because the number of degrees of freedom and co-annihilation channels increase with weak isospin $j$, which accounts for the orders of the red, blue and black solid(dashed) lines in Fig.~\ref{pic:monojet_results}.

It is well known that the relic abundance is inversely proportional to the DM annihilation cross section. For a fixed weak isospin $j$, the the annihilation cross sections of the Dirac DMs are twice as many as those of Majorana DMs, therefore one naively expects to see smaller relic abundance of Dirac DMs. However, an opposite order is depicted in Fig.~\ref{pic:monojet_results}, e.g. the D1 (D2, D3) model exhibits larger relic abundances than the M1 (M2, M3) model, respectively. That is owing to the treatment of DM density as explained in Ref.~\cite{Gondolo:1990dk}. For the Boltzmann equation
\beq
\dot{n}+3Hn=-\ave{\sigma v}(n^2-n^2_{\rm eq})
\eeq
in the calculation of thermal relic, there is a factor of 1/2 in front of $\ave{\sigma v}$ for the Dirac case, and no extra factor for the Majorana case, which causes the orders of the Dirac and Majorana DMs shown in Fig.~\ref{pic:monojet_results}.

\begin{figure}
\includegraphics[scale=0.35]{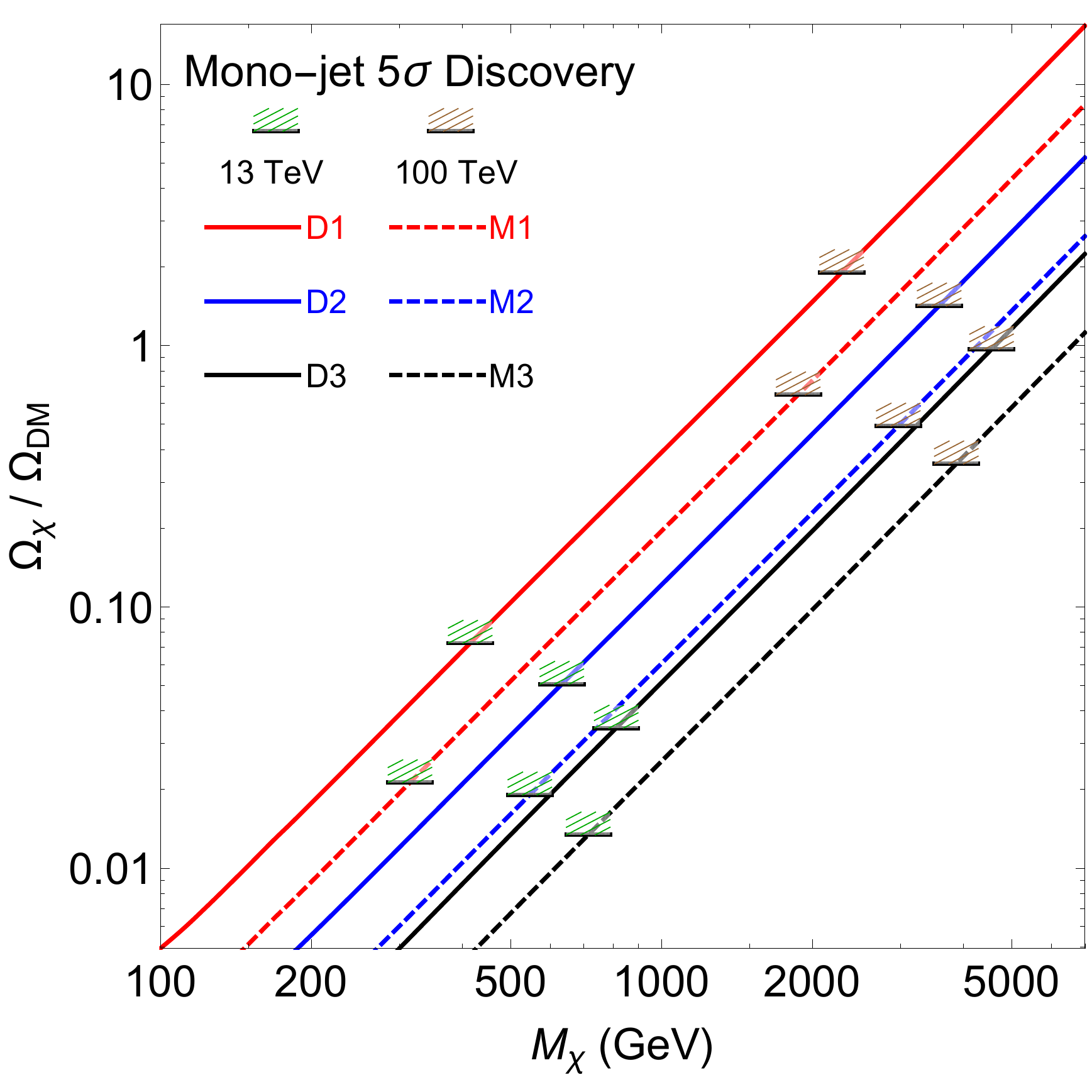}
\caption{The fraction $\mathcal{F}=\Omega_\chi/\Omega_{\rm DM}$ as a function of $m_{\chi_0}$ in the MDM models where the horizontal minibars label the upper limits of $m_{\chi_0}$ for claiming a $5\sigma$ discovery of DM in the mono-jet channel at the HL-LHC ($\mathcal{L}=3~{\rm ab}^{-1}$) and the 100~TeV collider ($\mathcal{L}=30~{\rm ab}^{-1}$).}
\label{pic:monojet_discovery}
\end{figure}

Following the same strategy, we consider the discovery potential of the MDM models at the both colliders. Figure~\ref{pic:monojet_discovery} displays the fraction $\mathcal{F}$ as a function of $m_{\chi_0}$ with horizontal minibars labelling the upper bounds on the DM mass for claiming a $5\sigma$ discovery in the mono-jet channel.  At the HL-LHC the DM needs to be very light as follows:
\bea
\text{D1:}&\quad& m_{\chi_0}\leqslant 415~{\rm GeV}, \quad \mathcal{F}\leqslant 7.3\%,\nn\\
\text{D2:}&\quad& m_{\chi_0}\leqslant 633~{\rm GeV}, \quad \mathcal{F}\leqslant 5.1\%,\nn\\
\text{D3:}&\quad& m_{\chi_0}\leqslant 813~{\rm GeV}, \quad \mathcal{F}\leqslant 3.4\%,\nn\\
\text{M1:}&\quad& m_{\chi_0}\leqslant 315~{\rm GeV}, \quad \mathcal{F}\leqslant 2.1\%,\nn\\
\text{M2:}&\quad& m_{\chi_0}\leqslant 546~{\rm GeV}, \quad \mathcal{F}\leqslant 1.9\%,\nn\\
\text{M3:}&\quad& m_{\chi_0}\leqslant 714~{\rm GeV}, \quad \mathcal{F}\leqslant 1.3\%.
\label{eq:monojet_discovery13}
\eea
Note that the parameter space of the D1 model is ruled out by the ATLAS result obtained in the disappearing-track signature~\cite{Aaboud:2017mpt}. If we do observe an excess in the mono-jet channel at the HL-LHC, then we reach the conclusion that none of the six MDM models considered in this work can explain the observed relic abundance and extra DM candidates are needed~\cite{Cao:2007fy}. 

The 100~TeV machine with $\mathcal{L}=30~{\rm ab}^{-1}$ is able to probe much heavier DMs as follows:
 \bea
\text{D1:}&\quad& m_{\chi_0}\leqslant 2289~{\rm GeV}, \quad \mathcal{F}\leqslant 190.2\%,\nn\\
\text{D2:}&\quad& m_{\chi_0}\leqslant 3584~{\rm GeV}, \quad \mathcal{F}\leqslant 141.7\%,\nn\\
\text{D3:}&\quad& m_{\chi_0}\leqslant 4558~{\rm GeV}, \quad \mathcal{F}\leqslant 96.7\%,\nn\\
\text{M1:}&\quad& m_{\chi_0}\leqslant 1875~{\rm GeV}, \quad \mathcal{F}\leqslant 64.8\%,\nn\\
\text{M2:}&\quad& m_{\chi_0}\leqslant 2970~{\rm GeV}, \quad \mathcal{F}\leqslant 49.2\%,\nn\\
\text{M3:}&\quad& m_{\chi_0}\leqslant 3877~{\rm GeV}, \quad \mathcal{F}\leqslant 35.2\%.
\label{eq:monojet_discovery100}
\eea
Of course, the mono-jet channel cannot discriminate the six MDM models, but it could tell whether or not the model can explain the observed relic abundance.  

\subsection{DM search in the mono-photon channel}

\begin{figure}[t]
\includegraphics[scale=0.6]{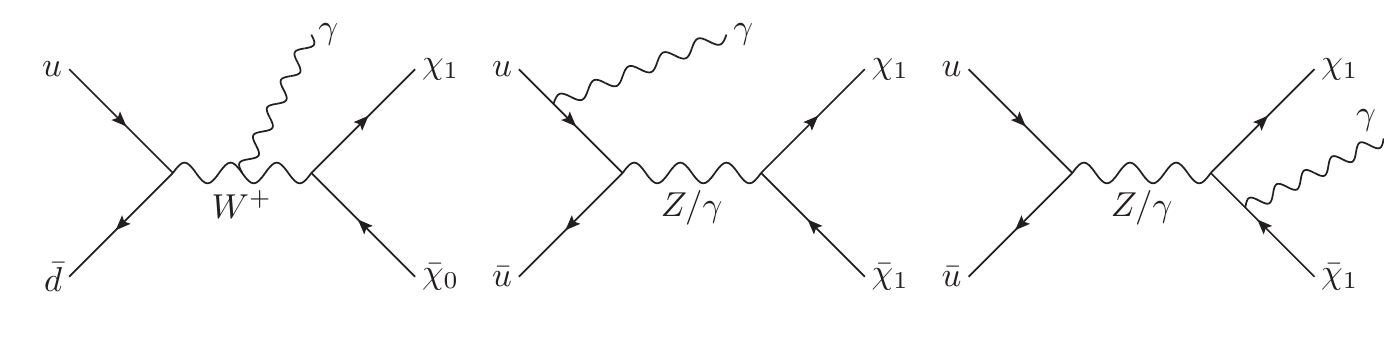}
\caption{Illustration of some Feynman diagrams for mono-photon channel in hadron collisions.}
\label{pic:monophoton}
\end{figure}

\begin{table}[b]
\caption{Exclusion limits of $m_{\chi_0}$ (in unit of GeV)  from the mono-photon searches at the 13 TeV LHC with $\mathcal{L}=3~{\rm ab}^{-1}$ (top) and at the 100 TeV collider with $\mathcal{L}=30~{\rm ab}^{-1}$ (bottom).}
\label{tab:monophoton}
\begin{tabular}{lcccccc}
\hline\hline
13 TeV &D1&D2&D3&M1&M2&M3\\
\hline
Cut 1  & 360 &  681 &  941 &  279  &  574  &  821   \\
Cut 2 & 361 &  696 &  965 &  275 &  586 &  840  \\
Cut 3 & 350 &  702 &  978 &  262 &  588 &  851  \\
\hline
The best & 361 &  702 &  978 &  279 &  588 &  851  \\
\hline
$\mathcal{F}=\Omega_\chi/\Omega_{\rm DM}$ &   5.5\%  &    6.2\%  &    4.9\%  &    1.7\%   &    2.2\%  &    1.9\%  \\
\hline\hline
~\\
\hline\hline
100 TeV &D1&D2&D3&M1&M2&M3\\
\colrule
Cut I & 1586&  3171&  4656 &  1220 &  2617 &  3958\\
Cut II & 1762 &  3460 &  5028 &  1384 &  2929 &  4333 \\
Cut III & 2017&  4022 &  5837 &  1543  &  3399&  4998 \\
\hline
The best & 2017 &  4022&  5837 &  1543 &  3399  &  4998  \\
\hline
$\mathcal{F}=\Omega_\chi/\Omega_{\rm DM}$ &   149.1\% &   177.4\%  &   156.8\%  &    44.5\%   &    63.9\%  &    57.9\%  \\
\hline\hline
\end{tabular}
\end{table}

Next we consider the mono-photon channel in which a pair of dark particles is produced in association with a hard photon. As depicted in Fig.~\ref{pic:monophoton}, the photon can be emitted from the quarks in the initial state, the $W^\pm$ boson in the intermediate state or the charged dark particles in the final state. The CMS collaboration has performed DM searches in the mono-photon channel at the 8~TeV and 13~TeV LHC~\cite{Khachatryan:2014rwa,Sirunyan:2017ewk}. The dominant backgrounds are the $Z(\to\nu\bar\nu)+\gamma$ and $W(\to\ell\nu)+\gamma$ $(\ell=e,\mu,\tau)$ processes. Other contributions, like $\gamma+$jets, $Z(\to\ell\bar\ell)+\gamma$, $t\bar{t}\gamma$, $VV\gamma$ and di-photon processes, are small and we discard these subdominant backgrounds from our analysis. To suppress the background events, at the 13~TeV (100~TeV) collider, we require the signal events consist of a hard photon in the central region of the detector and a large $\met$ from those unresolved dark particles in the final state. 
We also require the leading photon (the photon with the hardest $p_T$) and the missing transverse momentum do not overlap in the azimuthal plane: $\Delta\phi(\vec{p}_T^{\gamma},\vec{\slashed{E}}_T) >$ 2 radians. In addition, a separation in the azimuthal plane of $\Delta\phi(\vec{p}_T^j,\vec{\slashed{\rm E}}_T) >$ 0.5 radians between the missing transverse momentum direction and each of the four highest transverse momentum jets with $p_T^j >$ 30 GeV and $|\eta^j| <$ 5 is needed. This requirement significantly suppresses spurious backgrounds from mismeasured jets in the QCD multi-jet background~\cite{Sirunyan:2017jix}.  We then veto events that have electrons or muons with $p_T^{e(\mu)} >$ 10 GeV and separated from the leading photon by $\Delta R = \sqrt{(\Delta\eta)^2+(\Delta\phi)^2} >$ 0.5. 

\begin{figure}
\includegraphics[scale=0.35]{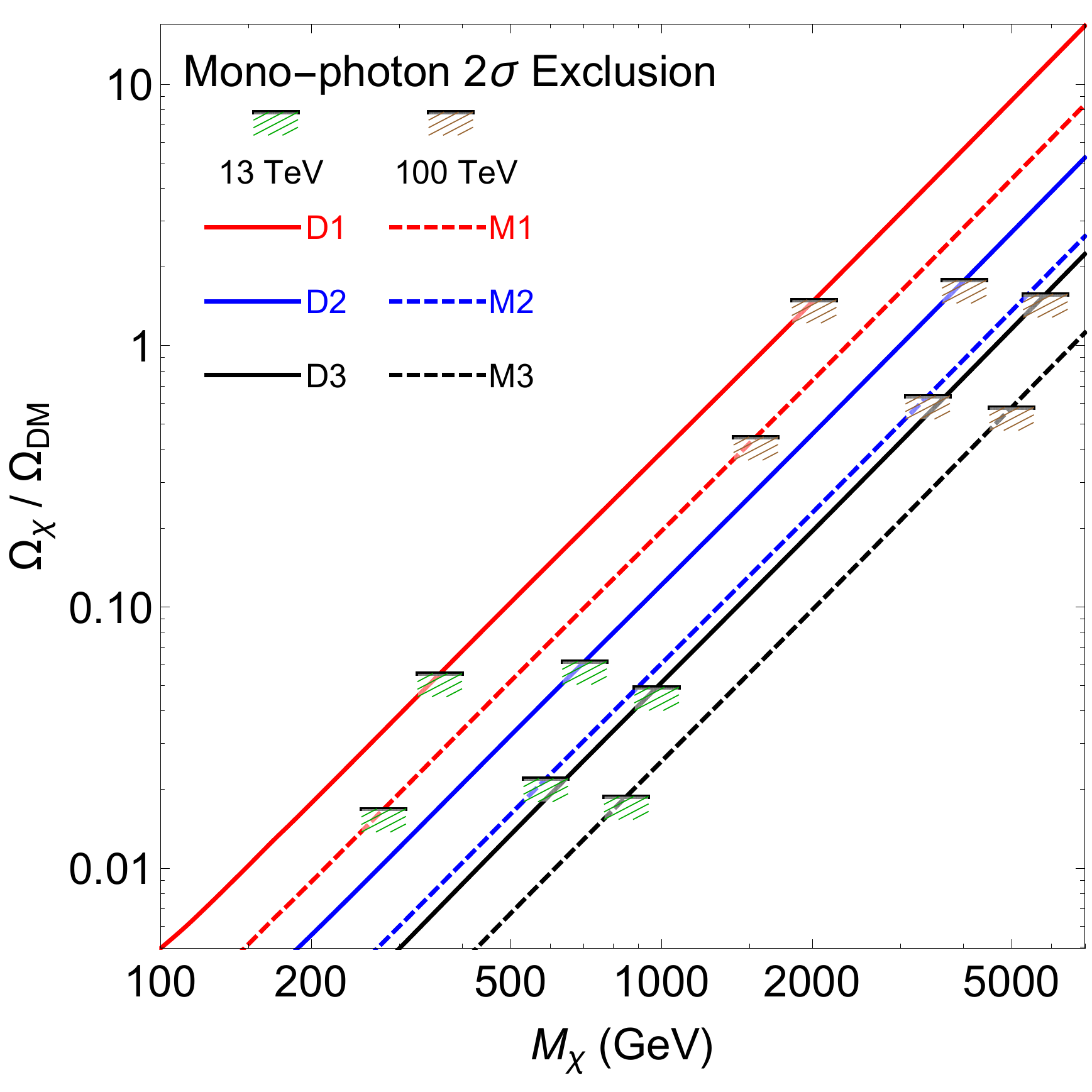}
\caption{The fraction $\mathcal{F}=\Omega_\chi/\Omega_{\rm DM}$ as a function of $m_{\chi_0}$ in the MDM models where the horizontal minibars label the lower limits of $m_{\chi_0}$ obtained from mono-photon channel at the HL-LHC ($\mathcal{L}=3~{\rm ab}^{-1}$) and the 100~TeV collider ($\mathcal{L}=30~{\rm ab}^{-1}$). 
}
\label{pic:monophoton_results}
\end{figure}

To further suppress the huge backgrounds, we impose hard cuts on $p_T^\gamma$ and $\met$. 
Three scenarios of hard cuts on $p_T^\gamma$ and $\met$ are considered in our study. At the 13~TeV LHC we follow Ref.~\cite{Sirunyan:2017ewk} to consider the hard cuts as follows:
\begin{align}
&&\text{cut-1}&:\quad & p_T^{\gamma}\geqslant175~{\rm GeV}, &&\met\geqslant170~{\rm GeV},\nn\\
&&\text{cut-2}&:\quad& p_T^{\gamma}\geqslant175~{\rm GeV}, &&\met\geqslant250~{\rm GeV},\nn\\
&&\text{cut-3}&:\quad & p_T^{\gamma}\geqslant330~{\rm GeV},&& \met\geqslant330~{\rm GeV}.\,
\label{eq:cut13gamma}
\end{align}
and increase the cut thresholds at the 100~TeV collider:
\begin{align}
&&\text{cut-I}&:\quad  p_T^{\gamma}\geqslant400~{\rm GeV}, && \met\geqslant200~{\rm GeV},\nn\\
&&\text{cut-II}&:\quad p_T^{\gamma}\geqslant400~{\rm GeV}, && \met\geqslant500~{\rm GeV},\nn\\
&&\text{cut-III}&:\quad p_T^{\gamma}\geqslant400~{\rm GeV}, &&\met\geqslant1000~{\rm GeV}.
\label{eq:cut100gamma}
\end{align}
Similar to the study of the mono-jet channel, we impose the optimal cuts shown above and use the best cut for each individual MDM model to obtain the best discovery or exclusion limit of $m_{\chi_0}$. Table~\ref{tab:monophoton} shows the exclusion limits of $m_{\chi_0}$ obtained from the three scenarios of optimal cuts at the 13~TeV LHC with an integrated luminosity of $3~{\rm ab}^{-1}$ (top panel) and at the 100~TeV collider with $\mathcal{L}=30~{\rm ab}^{-1}$ (bottom panel). We pick up the best $2\sigma$ lower bounds of $m_{\chi_0}$ and translate them into the lower  bounds of the fraction $\mathcal{F}$. The results are also depicted in Fig.~\ref{pic:monophoton_results}.

 As the production rate of the mono-photon channel is smaller than the rate of the mono-jet channel, weaker bounds on $m_{\chi_0}$ and $\mathcal{F}$ are obtained from the mono-photon channel. At the 100~TeV collider, one can exclude the D1, D2 and D3 MDM models, assuming no extra DM candidates.

\begin{figure}
\includegraphics[scale=0.35]{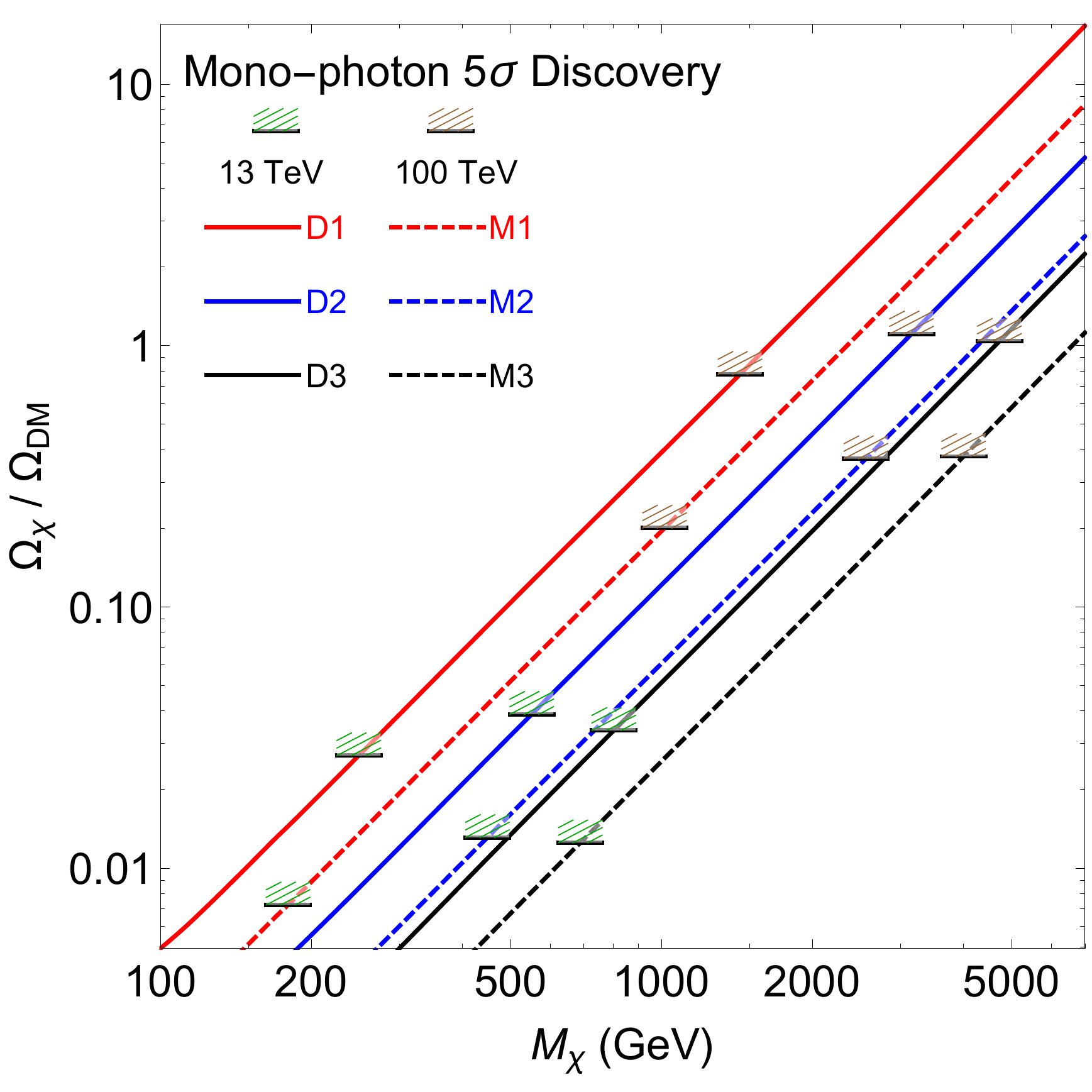}
\caption{The fraction $\mathcal{F}=\Omega_\chi/\Omega_{\rm DM}$ as a function of $m_{\chi_0}$ in the MDM models where the horizontal minibars label the upper limits of $m_{\chi_0}$ for claiming a $5\sigma$ discovery of DM in the mono-photon channel at the HL-LHC ($\mathcal{L}=3~{\rm ab}^{-1}$) and the 100~TeV collider ($\mathcal{L}=30~{\rm ab}^{-1}$).}
\label{pic:monophoton_discovery}
\end{figure}

We also consider the discovery potential of the MDM models at the HL-LHC and the 100~TeV collider ($\mathcal{L}=30~{\rm ab}^{-1}$). Figure~\ref{pic:monophoton_discovery} displays the fraction $\mathcal{F}$ as a function of $m_{\chi_0}$ with horizontal minibars labelling the upper bounds on the DM mass for claiming a $5\sigma$ discovery in the mono-photon channel.  At the HL-LHC the DM needs to be very light as follows:
\bea
\text{D1:}&\quad& m_{\chi_0}\leqslant 249~{\rm GeV}, \quad \mathcal{F}\leqslant 2.7\%,\nn\\
\text{D2:}&\quad& m_{\chi_0}\leqslant 551~{\rm GeV}, \quad \mathcal{F}\leqslant 3.9\%,\nn\\
\text{D3:}&\quad& m_{\chi_0}\leqslant 804~{\rm GeV}, \quad \mathcal{F}\leqslant 3.4\%,\nn\\
\text{M1:}&\quad& m_{\chi_0}\leqslant 180~{\rm GeV}, \quad \mathcal{F}\leqslant 0.7\%,\nn\\
\text{M2:}&\quad& m_{\chi_0}\leqslant 450~{\rm GeV}, \quad \mathcal{F}\leqslant 1.3\%,\nn\\
\text{M3:}&\quad& m_{\chi_0}\leqslant 689~{\rm GeV}, \quad \mathcal{F}\leqslant 1.2\%.
\label{eq:monojet_discovery13}
\eea
Taking account of the ATLAS search result in the disappearing-track signature~\cite{Aaboud:2017mpt}, the discovery parameter space of the D1 model is ruled out and the the discover parameter space of the D2 model is narrowed down to $460~{\rm GeV}\leqslant m_{\chi_0}\leqslant 551~{\rm GeV}$. The 100~TeV machine with $\mathcal{L}=30~{\rm ab}^{-1}$ is able to probe much heavier DMs as follows:
 \bea
\text{D1:}&\quad& m_{\chi_0}\leqslant 1435~{\rm GeV}, \quad \mathcal{F}\leqslant 77.5\%,\nn\\
\text{D2:}&\quad& m_{\chi_0}\leqslant 3155~{\rm GeV}, \quad \mathcal{F}\leqslant 110.6\%,\nn\\
\text{D3:}&\quad& m_{\chi_0}\leqslant 4727~{\rm GeV}, \quad \mathcal{F}\leqslant 103.8\%,\nn\\
\text{M1:}&\quad& m_{\chi_0}\leqslant 1015~{\rm GeV}, \quad \mathcal{F}\leqslant 20.0\%,\nn\\
\text{M2:}&\quad& m_{\chi_0}\leqslant 2559~{\rm GeV}, \quad \mathcal{F}\leqslant 36.9\%,\nn\\
\text{M3:}&\quad& m_{\chi_0}\leqslant 4009~{\rm GeV}, \quad \mathcal{F}\leqslant 37.6\%.
\label{eq:monojet_discovery100}
\eea
As expected, the performance of the mono-photon channel is worse than that of the mono-jet channel.

\subsection{DM search in the vector boson fusion channel}

\begin{figure}
\includegraphics[scale=0.36]{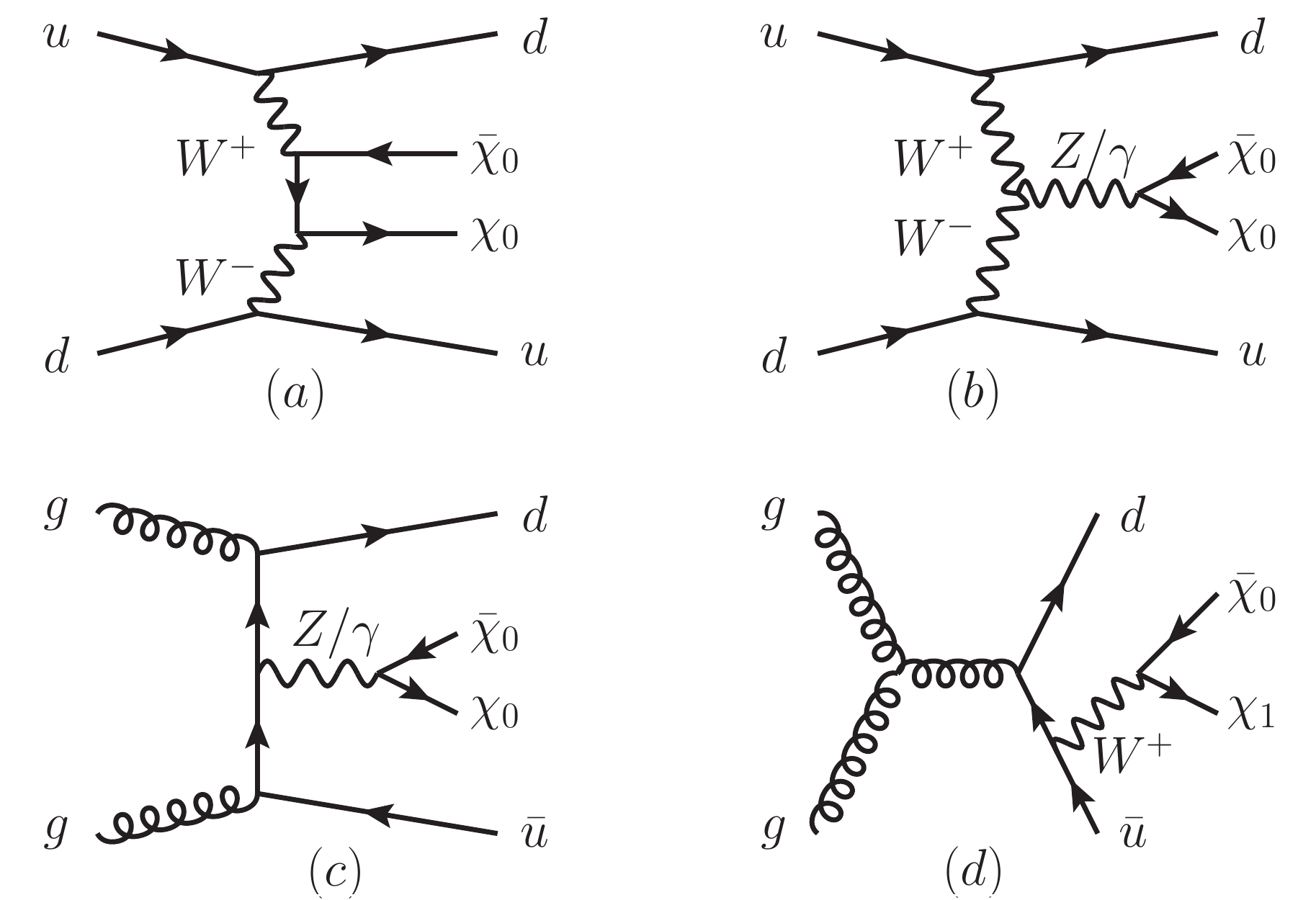}
\caption{Representative Feynman diagrams of VBF processes.}
\label{pic:vbf}
\end{figure}

Finally, we consider the VBF channel, which results in a collider signature of two hard jets in the forward region with large invariant mass plus a large missing transverse momentum. Figure~\ref{pic:vbf} shows a few representative Feynman diagrams. Notice that some diagrams not properly originating from two vector bosons (such as some QCD processes) also contribute to the signal and background events; see Fig.~\ref{pic:vbf}(c) and (d). Nevertheless we include those non-vector-boson diagrams in our study of the VBF channel.  
The CMS collaboration has searched such channel at the 8~TeV LHC~\cite{Khachatryan:2016mbu}. The dominant backgrounds are $Z(\to\nu\bar\nu)$+jets and $W^\pm(\to\ell^\pm\nu)$+jets $(\ell=e,\mu,\tau)$ processes. Other contributions, like $Z(\to\ell^+\ell^-)+$jets, $t\bar{t}$, single-top and di-boson processes, can be neglected after imposing cuts to be discussed below. 

To suppress the background events, we require exactly two jets with
\begin{align}
& p_T^{j_1(j_2)} >50~{\rm GeV}, && |\eta^{j_1(j_2)}|<5,\nn\\
& \eta^{j_1}\eta^{j_2}<0,&&  |\eta^{j_1}-\eta^{j_2}| > 4.2~,
\end{align}
and further demand hard cuts on the invariant mass of the two jets $m_{jj}$
and $\met$. In this study we consider eight scenarios of hard cuts which are summarized as follows. At the 13~TeV LHC,  we demand~\cite{Khachatryan:2016mbu}
\begin{align}
&\text{cut-1}&:\quad & m_{jj}\geqslant~750~{\rm GeV},  &&\met\geqslant250~{\rm GeV},\nn\\
&\text{cut-2}&:\quad & m_{jj}\geqslant1500~{\rm GeV}, && \met\geqslant250~{\rm GeV},\nn\\
&\text{cut-3}&:\quad & m_{jj}\geqslant1500~{\rm GeV}, && \met\geqslant400~{\rm GeV},\nn\\
&\text{cut-4}&:\quad & m_{jj}\geqslant2000~{\rm GeV}, && \met\geqslant350~{\rm GeV},\nn\\
&\text{cut-5}&:\quad & m_{jj}\geqslant2000~{\rm GeV}, && \met\geqslant400~{\rm GeV},\nn\\
&\text{cut-6}&:\quad & m_{jj}\geqslant2500~{\rm GeV}, && \met\geqslant350~{\rm GeV},\nn\\
&\text{cut-7}&:\quad & m_{jj}\geqslant2500~{\rm GeV}, && \met\geqslant400~{\rm GeV},\nn\\
&\text{cut-8}&:\quad & m_{jj}\geqslant3000~{\rm GeV}, && \met\geqslant400~{\rm GeV},\label{eq:cut13vbf}
\end{align}
and we raise the cut thresholds at the 100~TeV collider
\begin{align}
&\text{cut-I}&:\quad & m_{jj}\geqslant1000~{\rm GeV},  && \met\geqslant400~{\rm GeV},\nn\\
&\text{cut-II}&:\quad & m_{jj}\geqslant2000~{\rm GeV},&& \met\geqslant400~{\rm GeV},\nn\\
&\text{cut-III}&:\quad & m_{jj}\geqslant3000~{\rm GeV},&&\met\geqslant400~{\rm GeV},\nn\\
&\text{cut-IV}&:\quad & m_{jj}\geqslant4000~{\rm GeV},&& \met\geqslant400~{\rm GeV},\nn\\
&\text{cut-V}&:\quad & m_{jj}\geqslant5000~{\rm GeV},&&\met\geqslant600~{\rm GeV},\nn\\
&\text{cut-VI}&:\quad & m_{jj}\geqslant6000~{\rm GeV},&& \met\geqslant600~{\rm GeV},\nn\\
&\text{cut-VII}&:\quad & m_{jj}\geqslant7000~{\rm GeV},&& \met\geqslant600~{\rm GeV},\nn\\
&\text{cut-VIII}&:\quad & m_{jj}\geqslant8000~{\rm GeV},&& \met\geqslant600~{\rm GeV}.
\label{eq:cut100vbf}
\end{align}
In addition, a separation in the azimuthal plane of $\Delta\phi(\vec{p}_T^j,\met) >$ 0.5 radians between the missing transverse momentum direction and the sub-leading jet is also required. We also veto the background events with charged leptons, light flavor jets and $b$-tagged jets as follows: 
\begin{align}
&e^\pm(\mu^\pm) &:\qquad & p_T^{e(\mu)} >10~{\rm GeV}, &&|\eta^{e(\mu)}| < 2.5,\nn\\
&\tau^\pm &:\qquad & p_T^\tau >15~{\rm GeV},&& |\eta^\tau| <2.5,\nn\\
&\text{light jets} &:\qquad & p_T^j >30~{\rm GeV}, &&\nn\\
&b\text{-jet} &:\qquad & p_T^b >20~{\rm GeV},&& |\eta^b| <2.4~.
\end{align}

\begin{figure}
\includegraphics[scale=0.35]{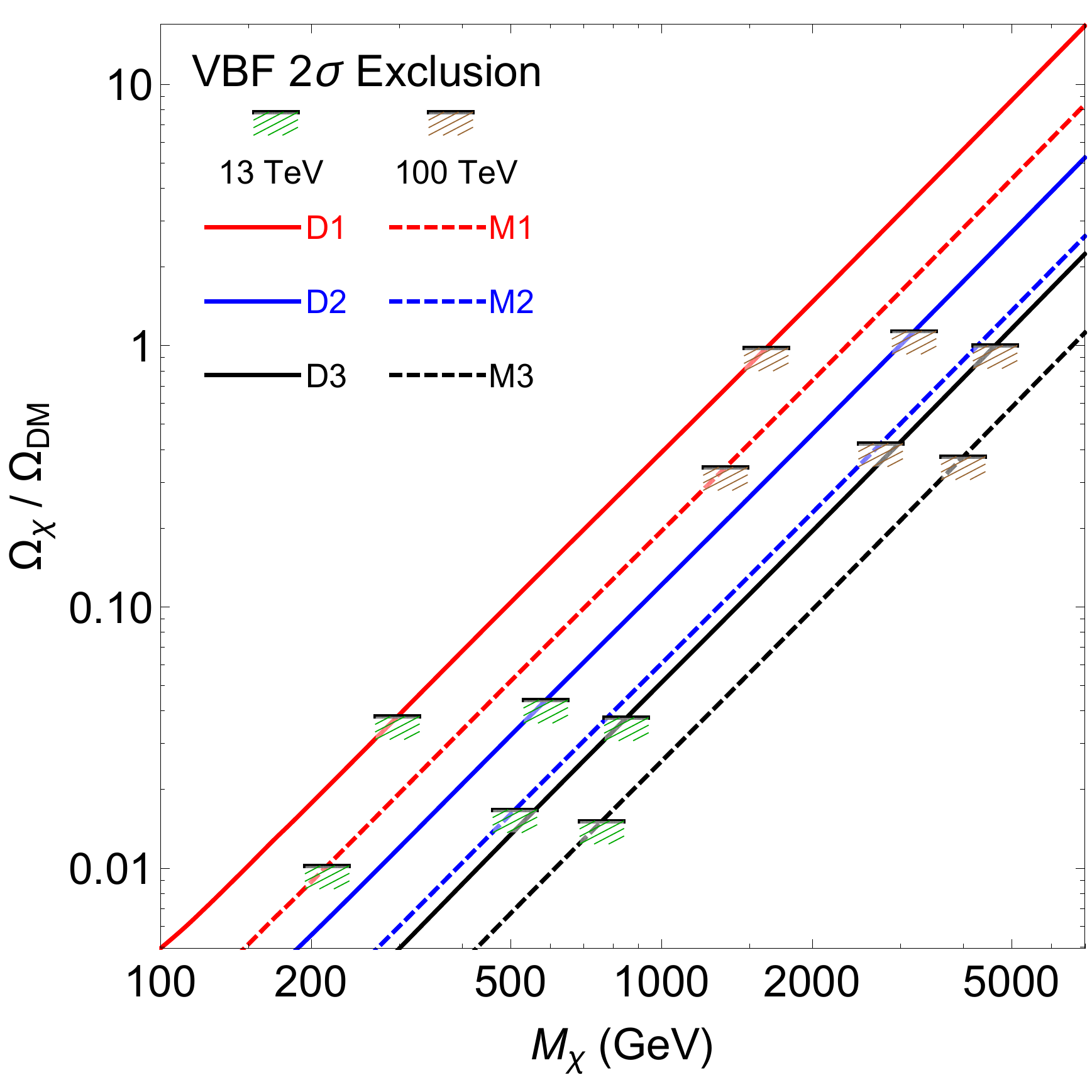}
\caption{The fraction $\mathcal{F}=\Omega_\chi/\Omega_{\rm DM}$ as a function of $m_{\chi_0}$ in the MDM models where the horizontal minibars label the upper limits of $m_{\chi_0}$ for claiming a $5\sigma$ discovery of DM in the VBF channel at the HL-LHC ($\mathcal{L}=3~{\rm ab}^{-1}$) and the 100~TeV collider ($\mathcal{L}=30~{\rm ab}^{-1}$).}
\label{pic:vbf_results}
\end{figure}

Table~\ref{tab:vbf_cms2} shows the exclusion limits of $m_{\chi_0}$ obtained from the eight scenarios of optimal cuts at the 13~TeV LHC with an integrated luminosity of $3~{\rm ab}^{-1}$ (top panel) and at the 100~TeV collider with $\mathcal{L}=30~{\rm ab}^{-1}$ (bottom panel). We pick up the best $2\sigma$ lower bounds of $m_{\chi_0}$ and translate them into the lower  bounds of the fraction $\mathcal{F}$. The results are also depicted in Fig.~\ref{pic:vbf_results}. The VBF channel turns out to be the weakest one to set relic abundance limits for the MDM models.

\begin{table}
\caption{Exclusion limits of $m_{\chi_0}$ (in unit of GeV) and $\mathcal{F}=\Omega_\chi/\Omega_{\rm DM}$ from the VBF channel at the HL-LHC with $\mathcal{L}=3~{\rm ab}^{-1}$ (top ) and at the 100~TeV collider with $\mathcal{L}=30~{\rm ab}^{-1}$ (bottom).}
\label{tab:vbf_cms2}
\begin{tabular}{lcccccc}
\hline\hline
13 TeV &D1&D2&D3&M1&M2&M3\\
\colrule
Cut-1  & 286  &  530  &  760    &  209    &  452   &  664  \\
Cut-2  & 288  &  546  &  777    &  214    &  471   &  679  \\
Cut-3  & 280  &  548  &  792    &  215    &  477   &  692  \\
Cut-4  & 275  &  546  &  781    &  200    &  461   &  683  \\
Cut-5  & 266  &  523  &  758    &  182    &  439   &  656  \\
Cut-6  & 295  &  583  &  845    &  198    &  508   &  747  \\
Cut-7  & 297  &  588  &  851    &  189    &  510   &  761  \\
Cut-8  & 271  &  565  &  829    &  168    &  496   &  702  \\
\hline
The best & 297  &  588   &  851    &  215    &  510 &  761 \\
\hline
$\mathcal{F}=\Omega_\chi/\Omega_{\rm DM}$ &   3.8\%  &    4.4\%  &    3.8\%   &    1.0\%   &    1.7\%  &    1.5\%  \\
\hline\hline
~\\
\hline\hline
100 TeV&D1&D2&D3&M1&M2&M3\\
\colrule
Cut-I        & 1265  &  2510  &  3635 & 1057  &  2106 &  3121 \\
Cut-II       & 1275  &  2553  &  3689 & 1079  &  2138 &  3187 \\
Cut-III      & 1275  &  2586  &  3788 & 1068  &  2171 &  3263 \\
Cut-IV       & 1221  &  2532  &  3711 & 1013  &  2084 &  3209 \\
Cut-V        & 1608  &  3193  &  4634 & 1342  &  2739 &  4002 \\
Cut-VI       & 1619  &  3182  &  4634 & 1320  &  2728 &  3991 \\
Cut-VII      & 1608  &  3182  &  4634 & 1198  &  2717 &  3991 \\
Cut-VIII     & 1597  &  3171  &  4612 & 1176  &  2706 &  3980 \\
\hline
The best     & 1619  &  3193  &  4634 & 1342  &  2739 &  4002 \\
\hline
$\mathcal{F}=\Omega_\chi/\Omega_{\rm DM}$ &   97.7\%  &   113.3\%  &    99.8\%  &   34.1\%  &    42.0\%  &    37.5\%  \\
\hline\hline
\end{tabular}
\end{table}

\begin{figure}[b]
\includegraphics[scale=0.35]{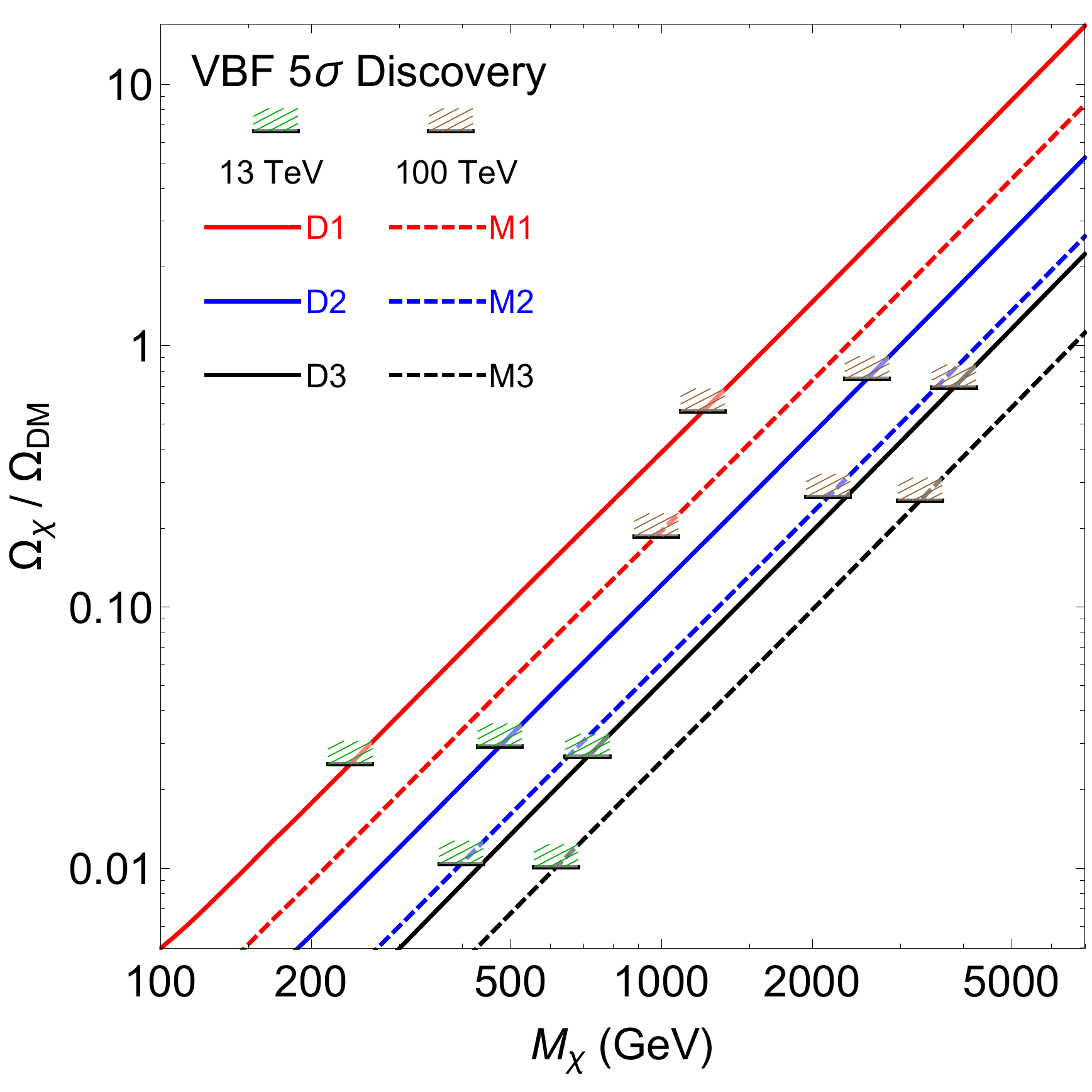}
\caption{The fraction $\mathcal{F}=\Omega_\chi/\Omega_{\rm DM}$ as a function of $m_{\chi_0}$ in the MDM models where the horizontal minibars label the upper limits of $m_{\chi_0}$ for claiming a $5\sigma$ discovery of DM in the VBF channel at the HL-LHC ($\mathcal{L}=3~{\rm ab}^{-1}$) and the 100~TeV collider ($\mathcal{L}=30~{\rm ab}^{-1}$).}
\label{pic:vbf_discovery}
\end{figure}

\begin{figure*}
\includegraphics[scale=0.4]{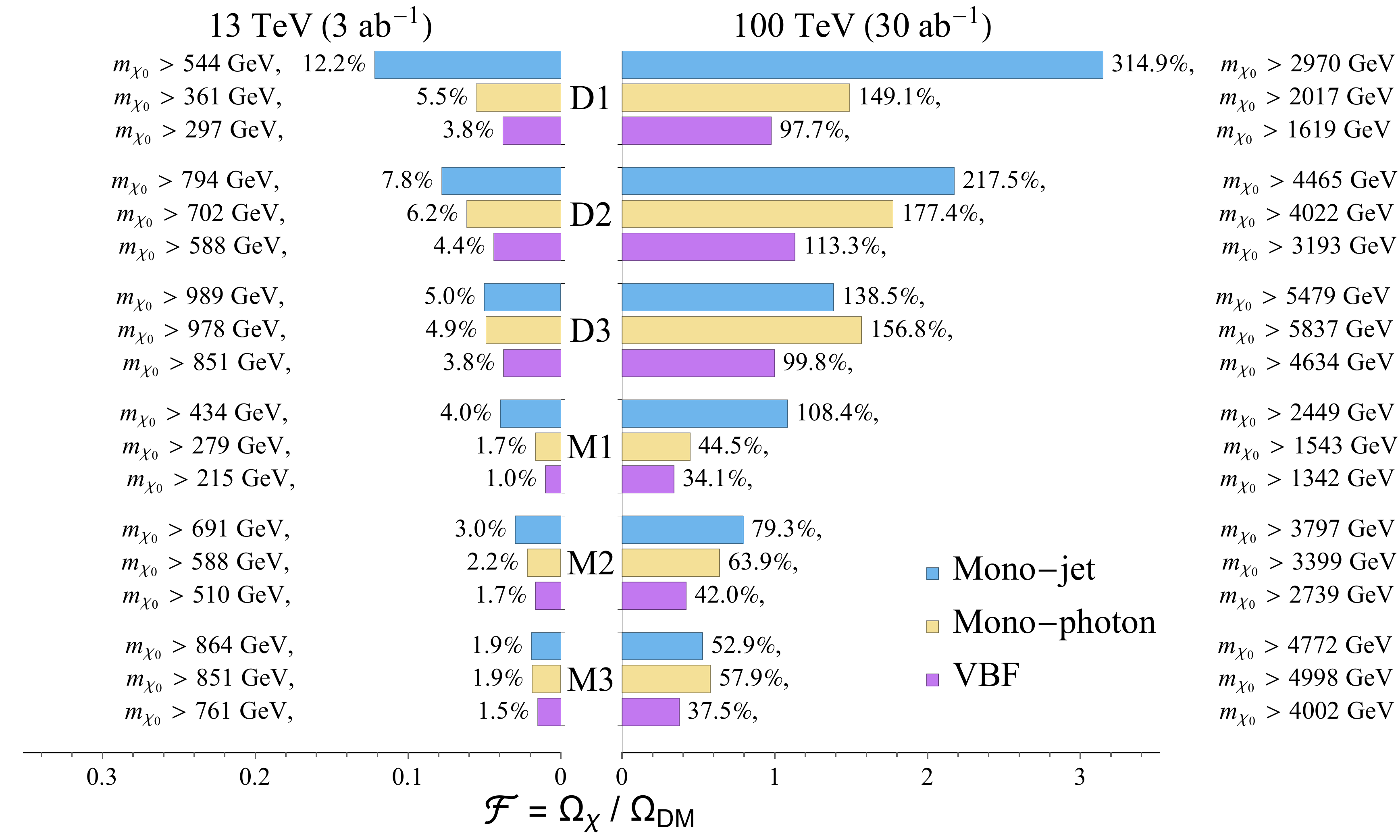}
\caption{The 95\% lower bounds on the fraction $\mathcal{F}=\Omega_\chi/\Omega_{\rm DM}$ in various MDM models where the shaded bands are ruled out by the mono-jet channel (blue), the mono-photon channel (orange) and the VBF channel (purple) at the HL-LHC (left) and the 30 ab$^{-1}$ 100~TeV collider (right). }
\label{fig:all_results}
\end{figure*}

Figure~\ref{pic:vbf_discovery} displays the fraction $\mathcal{F}$ as a function of $m_{\chi_0}$ with horizontal minibars labelling the upper bounds on the DM mass for claiming a $5\sigma$ discovery in the mono-photon channel.  At the HL-LHC the DM needs to be very light as follows:
\bea
\text{D1:}&\quad& m_{\chi_0}\leqslant 240~{\rm GeV}, \quad \mathcal{F}\leqslant 2.5\%,\nn\\
\text{D2:}&\quad& m_{\chi_0}\leqslant 476~{\rm GeV}, \quad \mathcal{F}\leqslant 2.9\%,\nn\\
\text{D3:}&\quad& m_{\chi_0}\leqslant 712~{\rm GeV}, \quad \mathcal{F}\leqslant 2.7\%,\nn\\
\text{M1:}&\quad& m_{\chi_0}\leqslant 104~{\rm GeV}, \quad \mathcal{F}\leqslant 0.3\%,\nn\\
\text{M2:}&\quad& m_{\chi_0}\leqslant 399~{\rm GeV}, \quad \mathcal{F}\leqslant 1.0\%,\nn\\
\text{M3:}&\quad& m_{\chi_0}\leqslant 616~{\rm GeV}, \quad \mathcal{F}\leqslant 1.0\%.
\label{eq:monojet_discovery13}
\eea
Taking account of the ATLAS search result in the disappearing-track signature~\cite{Aaboud:2017mpt}, the discovery parameter space of the D1 model is ruled out and the the discover parameter space of the D2 model is narrowed down to $460~{\rm GeV}\leqslant m_{\chi_0}\leqslant 476~{\rm GeV}$.

The 100~TeV machine with $\mathcal{L}=30~{\rm ab}^{-1}$ is able to probe much heavier DMs as follows:
 \bea
\text{D1:}&\quad& m_{\chi_0}\leqslant 1209~{\rm GeV}, \quad \mathcal{F}\leqslant 55.9\%,\nn\\
\text{D2:}&\quad& m_{\chi_0}\leqslant 2573~{\rm GeV}, \quad \mathcal{F}\leqslant 74.5\%,\nn\\
\text{D3:}&\quad& m_{\chi_0}\leqslant 3836~{\rm GeV}, \quad \mathcal{F}\leqslant 69.0\%,\nn\\
\text{M1:}&\quad& m_{\chi_0}\leqslant ~977~{\rm GeV}, \quad \mathcal{F}\leqslant 18.6\%,\nn\\
\text{M2:}&\quad& m_{\chi_0}\leqslant 2151~{\rm GeV}, \quad \mathcal{F}\leqslant 26.3\%,\nn\\
\text{M3:}&\quad& m_{\chi_0}\leqslant 3282~{\rm GeV}, \quad \mathcal{F}\leqslant 25.5\%.
\label{eq:monojet_discovery100}
\eea
Again, the VBF channel is the weakest channel to discover the DM in the MDM models.

\section{Summary and outlook}\label{sec:summary}

Recent DM direct searches place very stringent constraints on the possible DM candidates proposed in extensions of the SM. In this work we consider the special case that the DM candidate escapes the DM direct-detection when the experimental sensitivity reaches the irreducible solar neutrino flux background. In such a bad circumstance one has to rely on the collider searches and indirect-detection programs to probe the DM. In this work we consider the minimal dark-matter model~\cite{Cirelli:2014dsa} and demonstrate that, thanks to the very few parameters in the model, the searching results of DM at the colliders are strongly correlated with the DM relic abundance in the early Universe.

We consider both the Dirac and Majorana fermion DM candidates in the weak triplet ($j=1$), quintet ($j=2$) and septet ($j=3$) representations, named as the D1, D2, D3, M1, M2 and M3 model, respectively. To avoid the DM direct detection, all the DM multiplets exhibit a  hypercharge $Y=0$. The EW loop corrections generate small mass splittings (about several hundred MeV) among the dark particles in the same weak representation and make the neutral dark particle as the lightest particle so as to be the DM candidate. Three different searching channels, i.e the mono-jet, mono-photon and VBF channels, are explored at the 13~TeV LHC with an integrated luminosity of $3~{\rm ab}^{-1}$ (HL-LHC) and at the 100~TeV collider with an integrated luminosity of $30~{\rm ab}^{-1}$.  
We follow the searching strategies used by the ATLAS and CMS collaborations~\cite{Sirunyan:2017jix,Sirunyan:2017ewk,Khachatryan:2016mbu} and find a good agreement. For simplicity we only present the results of using the CMS strategy in this work. We first obtain the 95\% C.L. bounds on the DM mass $m_{\chi_0}$ if no excesses are observed in the mono-jet, mono-photon and VBF channels. It is shown that using the disappearing tracks is more sensitive to DM searches than the conventional mono-$X$ method~\cite{Cirelli:2014dsa,Han:2018wus}, but both the mono-$X$ and disappearing track methods will provide  complementary informations. 

The mass constraints derived from the collider searches can be translated into the lower limits of the DM relic abundance. Considering the cosmological observable value $\Omega_{\rm DM} h^2\approx0.12$, we use the fraction $\mathcal{F}$($\equiv\Omega_\chi/\Omega_{\rm DM}$) to describe the quota of the MDM inside $\Omega_{\rm DM}$. Figure~\ref{fig:all_results} displays the fraction $\mathcal{F}$'s derived from the mono-jet channel (blue), the mono-photon channel (orange) and the VBF channel (purple) at the HL-LHC (left) and the 100~TeV collider (right). The colliders are sensitive to the Dirac-type DMs more than to the Majorana DMs. For example, the exclusion fractions of the Dirac DMs is roughly about $2.5\sim 3$ times larger than those of the Majorana DMs.  We also note that the relic abundance constrains at the 30 ab$^{-1}$ 100 TeV collider are about $25\sim 30$ times larger than those constraints at the HL-LHC,  indicating that the future high-energy and high-luminosity hadron colliders have a considerable enhancement to bound the relic abundance for the MDM models. In particular, if we assume one and only one DM candidate, the D1, D2, D3 and M1 models are excluded at the 95\% C.L. at the 100~TeV collider as they predict too much DM relic abundance.

In conclusion, the HL-LHC and the future 100 TeV colliders can be used to measure relic abundance of DM, not only for the MDM model, but also for other WIMP models in which the relic abundance and collider searches are highly correlated.  The results show that the future high-energy collider could considerably extend the constraints for the MDM relic abundance. The correlation will give us some enlightening connections between particle physics and cosmology. 

\begin{acknowledgments}
The work is supported in part by the National Science Foundation of China under Grant Nos. 11275009, 11675002, 11635001 and 11725520. KPX is supported by the National Research Foundation of Korea under grant 2017R1D1A1B03030820. 
\end{acknowledgments}

\appendix

\section{The MDM model}\label{app:MDM}
In the appendix we briefly introduce the minimal dark matter models used in our analysis. 

\subsection{Dirac MDM model}

Let $\chi=\begin{pmatrix}\chi_j&\chi_{j-1}&...&\chi_{-j}\end{pmatrix}^T$ be an $SU(2)_L$ multiplet with weak isospin $j$. In order to couple with SM gauge bosons, we must have $j\neq0$. Some papers make additional assumptions to the case $j=0$, for example, a heavy dark vector boson $X_\mu$ as a mediator is introduced in~\cite{Chua:2013zpa}. But since this additional assumption is not ``minimal'', we will not consider it here, i.e., we have $j\geqslant1$, $j\in\mathbb{Z}$ in this paper. The Lagrangian is then simply
$$
\mathcal{L}_\chi^{\rm Dirac}=\bar\chi i\slashed{D}\chi-M_\chi\bar\chi\chi,
$$
where $D_\mu=\partial_\mu-igT^iA_\mu^i-ig'Y_\chi B_\mu$ is the gauge covariant derivative, and $T^i$ are matrix representations of the three generators of the $SU(2)_L$ group. Working in the representation that $T^3$ is diagonal, with the indices noted as
\beq
T_{m_1m_2}=\begin{pmatrix}T_{j,j}&T_{j,j-1},&\cdots&T_{j,-j}\\
T_{j-1,j}&T_{j-1,j-1}&\cdots&T_{j-1,-j}\\
\vdots&\vdots&\ddots&\vdots\\
T_{-j,j}&T_{-j,j-1}&\cdots&T_{-j,-j}\end{pmatrix},
\eeq
the elements of $T$ matrices can be expressed as
\bea
T^+_{m_1m_2}&=&\delta_{m_1-1,m_2}\sqrt{(j+m_1)(j-m_1+1)},\nn\\
T^-_{m_1m_2}&=&\delta_{m_2-1,m_1}\sqrt{(j+m_2)(j-m_2+1)},\nn\\
T^3_{m_1m_2}&=&\delta_{m_1,m_2}m_1,
\eea
where repeated indices do not mean summation. Expanding the generators gives rise to Eq.~(\ref{eq:lag1}).

\subsection{Majorana MDM model}

The most convenient tool to deal with Majorana field is the 2-component Weyl spinor and dot notation. Consider a multiplet Weyl spinor $\xi=\begin{pmatrix}\xi_j&\xi_{j-1}&...&\xi_{-j}\end{pmatrix}^T$ with weak isospin $j$. To write down a mass term we must set $Y_\xi=0$, and then the Lagrangian reads as
\bea
\mathcal{L}_\xi^{\rm Majorana}&=&\xi^\dagger i\bar\sigma^\mu D_\mu\xi-\frac{M_\chi}{2}(\xi U \xi+{\rm h.c.})\nn\\
&=&\xi^\dagger\bar\sigma^\mu\Big[i\partial_\mu+\frac{g}{\sqrt{2}}(T^-W^-_\mu+T^+W^+_\mu)\nn\\
&&\qquad~~ +gc_WT^3Z_\mu+gs_WT^3A_\mu\Big]\xi\nn\\
&-&\frac{M_\chi}{2}(\xi U \xi+{\rm h.c.}),
\eea
where $U$ is the CG coefficient matrix
\beq
\xi U\xi\propto\sum_{m=j}^{-j}\frac{(-1)^{j-m}}{\sqrt{2j+1}}\xi_m\xi_{-m},
\eeq
with
\beq
\frac{(-1)^{j-m}}{\sqrt{2j+1}}=\langle j,m;j,-m|0,0\rangle.
\eeq
For normalization we choose
\beq
\xi U\xi=\sum_{m=j}^{-j}(-1)^{m}\xi_m\xi_{-m}.
\eeq
Expanding the generator matrices $T^i$, we have
\begin{widetext}
\bea
\mathcal{L}_\xi&=&\sum_{m=j}^{-j}\xi^\dagger_mi\bar\sigma^\mu\partial_\mu\xi_m-\frac{M_\chi}{2}\sum_{m=j}^{-j}(-1)^{m}(\xi_m\xi_{-m}+{\rm h.c.})+\sum_{m=j}^{-j}mg(c_WZ_\mu+s_WA_\mu)\xi^\dagger_m\bar\sigma^\mu\xi_m\nn\\
&+&\sum_{m=j-1}^{-j}\sqrt{(j+m+1)(j-m)}\frac{g}{\sqrt{2}}(W^-_\mu\xi^\dagger_m\bar\sigma^\mu\xi_{m+1}+W^+_\mu\xi^\dagger_{m+1}\bar\sigma^\mu\xi_m)\nn\\
&=&\xi^\dagger_0i\bar\sigma^\mu\partial_\mu\xi_0-\frac{M_\chi}{2}(\xi_0\xi_0+{\rm h.c.})+\sum_{m=1}^{j}(\xi^\dagger_mi\bar\sigma^\mu\partial_\mu\xi_m+\xi^\dagger_{-m}i\bar\sigma^\mu\partial_\mu\xi_{-m})-M_\chi\sum_{m=1}^{j}(-1)^{m}(\xi_m\xi_{-m}+{\rm h.c.})\nn\\
&&+\sum_{m=1}^{j}mg(c_WZ_\mu+s_WA_\mu)(\xi^\dagger_m\bar\sigma^\mu\xi_m-\xi^\dagger_{-m}\bar\sigma^\mu\xi_{-m})\nn\\
&&+\left(\sum_{m=1}^{j}\sqrt{(j+m)(j-m+1)}\frac{g}{\sqrt{2}}W^-_\mu(\xi^\dagger_{m-1}\bar\sigma^\mu\xi_m+\xi^\dagger_{-m}\bar\sigma^\mu\xi_{-m+1})+{\rm h.c.}\right).
\eea
\end{widetext}
We thus obtain Eq.~(\ref{eq:lag2}) after defining
\bea
&& \chi_0=\begin{pmatrix}\xi_{0\alpha}\\ \xi^{\dagger\dot\alpha}_0\end{pmatrix},\nn\\
&&\chi_m=\begin{pmatrix}\xi_{m\alpha}\\(-1)^m \xi^{\dagger\dot\alpha}_{-m}\end{pmatrix},\quad m=1,~2,~\cdots,~j.
\eea

\section{Annihilation cross sections of Dirac-type DM }\label{sec:xsection-Dirac}

We summarize the cross sections of dark particles annihilating into the SM particles in the MDM models. The non-relativistic expansion coefficients $a$ and $b$ of each annilation channel are also shown. As the DM mass of interest to us is very large, we treat all the SM particles as massless to simplify the analytical expression. We further ignore the tiny mass splittings between dark particles. 

Below we only present the processes with zero or positive electric charge in the initial/final state, such as $\chi_0 \bar{\chi}_{-1} \to \nu_e e^+$ in the Dirac model. The corresponding conjugate process, i.e., $\chi_{-1} \bar{\chi}_0 \to e^- \bar{\nu}_e$, is omitted can be easily obtained. We introduce shorthand notations as follows: 
\begin{align}
&\ell=\{e,\mu,\tau\}, && \nu_\ell=\{\nu_e,\nu_\mu,\nu_\tau\}, && {\rm L}=\{e,\mu,\tau,\nu_e,\nu_\mu,\nu_\tau\}\nn\\
&q_u=\{u,c,t\},  &&q_d=\{d,s,b\}, && {\rm Q}=\{u,c,t,d,s,b\},
\end{align}
and we choose the CKM matrix elements as diagonal for simplicity. 
Defining $\beta\equiv 4M_\chi^2/s$, we introduce two terms that often shows up in the formulas: 
\beq
{\rm LT}=\log \left(\frac{1+\sqrt{1-\beta}}{1-\sqrt{1-\beta}}\right), \quad {\rm ST}=\sqrt{1-\beta}.
\eeq
Next we present the DM annihilation cross section in accord to the SM particles in the final state. The symbol ``$j$" denotes the weak isospin of dark particles, the symbol ``$m$" is the $T^3$ eigenvalue of the dark particle $\chi_m$, and $N_c=3$ is the color factor of quarks. 
We further define a few functions as follows: 
\begin{widetext}
\begin{eqnarray}
&& F_1(s,M_\chi) =\frac{\left(2 M_{\chi }^2+s\right)}{s^{3/2} \sqrt{s-4 M_{\chi }^2}},\nn\\
&& F_2(s,M_\chi)=\frac{ {\rm LT} \left(4 s M_{\chi }^2-8 M_{\chi }^4+s^2\right)-{\rm ST} \left(s^2+4 sM_{\chi }^2\right)}{s^2 \left(s-4 M_{\chi }^2\right)},\nn\\
&& F_3(s,M_\chi)=\frac{3 {\rm LT} \left(s^2+4 s M_{\chi }^2\right)-{\rm ST} \left(5s^2+22 sM_{\chi }^2 \right)}{ s^2 \left(s-4 M_{\chi }^2\right)},\nn\\
&& F_4(s,M_\chi)=\frac{3 {\rm LT} \left(20 s M_{\chi }^2-32 M_{\chi }^4+5 s^2\right)-{\rm ST} \left(17 s^2+70 sM_{\chi }^2\right)}{ s^2 \left(s-4 M_{\chi }^2\right)},\nn\\
&& F_5(s,M_\chi)=\frac{3 {\rm LT} \left(52 s M_{\chi }^2-96 M_{\chi }^4+13 s^2\right)-{\rm ST} \left(41 s^2+166 sM_{\chi }^2\right)}{s^2 \left(s-4 M_{\chi }^2\right)},\nn\\
&&F_6(s,M_\chi)=\frac{24 {\rm LT} \left(g^2-e^2\right) \left(4 s M_{\chi }^2+s^2\right)-{\rm ST} \left(\left(174 g^2-176 e^2\right) sM_{\chi }^2+s^2 \left(39 g^2-40 e^2\right)\right)}{ s^2 \left(s-4 M_{\chi }^2\right)},\nn\\
&&F_7(s,M_\chi)=\frac{24 {\rm LT} \left(g^2-e^2\right) \left(20 s M_{\chi }^2-32 M_{\chi }^4+5 s^2\right)-{\rm ST} \left(\left(558 g^2-560 e^2\right) sM_{\chi }^2+s^2 \left(135 g^2-136 e^2\right)\right)}{s^2 \left(s-4 M_{\chi }^2\right)},\nn\\
&& F_8(s,M_\chi)=\frac{24 {\rm LT} \left(g^2-e^2\right) \left(52 s M_{\chi }^2-96 M_{\chi }^4+13 s^2\right)-{\rm ST} \left(\left(1326 g^2-1328 e^2\right) sM_{\chi }^2+s^2 \left(327 g^2-328 e^2\right)\right)}{s^2 \left(s-4 M_{\chi }^2\right)},\nn\\
&& F_9(s,M_\chi)=\frac{8 {\rm LT} \left(4 s M_{\chi }^2+s^2\right)-{\rm ST} \left(13 s^2+58 s M_{\chi }^2\right)}{ s^2 \left(s-4 M_{\chi }^2\right)},\nn\\
&& F_{10}(s,M_\chi)=\frac{8 {\rm LT} \left(52 s M_{\chi }^2-96 M_{\chi }^4+13 s^2\right)-{\rm ST} \left(109 s^2 + 442 s M_{\chi }^2\right)}{ s^2 \left(s-4 M_{\chi }^2\right)},\nn\\
&& F_{11}(s,M_\chi)=\frac{8 {\rm LT} \left(244 s M_{\chi }^2-480 M_{\chi }^4+61 s^2\right)-{\rm ST} \left(493 s^2 +1978 s M_{\chi }^2\right)}{s^2 \left(s-4 M_{\chi }^2\right)},\nn\\
&& F_{12}(s,M_\chi)=\frac{4 {\rm LT} \left(68 s M_{\chi }^2-120 M_{\chi }^4+17 s^2\right)-{\rm ST} \left(73 s^2+298 s M_{\chi }^2 \right)}{s^2 \left(s-4 M_{\chi }^2\right)}.
\end{eqnarray}
\end{widetext}

\subsection{Fermion pairs}

The annihilation mode of a pair of SM fermions can only occur through the $s$-channel diagram mediated by the SM gauge bosons, which yields the cross section and coefficients as follows: 
\begin{itemize}[leftmargin=*]
\item $\chi_m\bar{\chi}_{m-1} \to \nu_\ell \ell^+$~:
\bea
&& g_{CG}=(j+m)(j-m+1),\nn\\
&&\sigma=g_{CG}\frac{g^4}{96\pi} F_1(s,M_\chi),\nn\\
&& a = g_{CG}\frac{g^4}{128\pi M_\chi^2},\quad b = -g_{CG}\frac{5g^4}{3072\pi M_\chi^2};
\eea
\item $\chi_m \bar{\chi}_m \to {\rm L} \overline{{\rm L}}$~:
\begin{align}
& g_{CG}=m^2,\nn\\
&\sigma=g_{CG}\frac{g^4}{96\pi} F_1(s,M_\chi),\nn\\
& a=g_{CG}\frac{g^4}{128 \pi  M_{\chi }^2}, \quad b=-g_{CG}\frac{5 g^4}{3072 \pi  M_{\chi }^2};
\end{align}
\item $\chi_m \bar{\chi}_{m-1} \to q_u \bar{q_d}$~:
 \bea
 && g_{CG}=(j+m)(j-m+1),\nn\\
&&\sigma=g_{CG}N_c\frac{g^4}{96\pi} F_1(s,M_\chi),\nn\\
&& a = g_{CG}N_c\frac{g^4}{128\pi M_\chi^2},\quad b = -g_{CG}N_c\frac{5g^4}{3072\pi M_\chi^2};
\eea 
\item $\chi_m \bar{\chi}_m \to {\rm Q} \bar{{\rm Q}}$~:
\begin{align}
& g_{CG}=m^2,\nn\\
&\sigma=g_{CG}N_c\frac{g^4}{96\pi} F_1(s,M_\chi),\nn\\
&a=g_{CG}N_c\frac{g^4}{128 \pi  M_{\chi }^2}, \quad b=-g_{CG}N_c\frac{5g^4}{3072 \pi  M_{\chi }^2}. 
\end{align}
\end{itemize}

\subsection{Gauge boson pairs}

Now consider the annihilation mode of a pair of gauge bosons which is more complicated than the fermion mode. Most of the processes occur through the $t$-channel diagram but a few of them can also be through the $s$-channel diagram. We thus separate the cross section expression depending on the topology of the annihilation diagrams.  

We first consider those processes occurring only through the $t$-channel diagram whose cross section and coefficients are presented as follows:
\begin{itemize}[leftmargin=*]
\item $\chi_0 \bar{\chi}_0 \to W^+ W^-$~:
\bea
&& g_{CG}=(j+m)^2(j-m+1)^2,\nn\\
&&\sigma =g_{CG} \frac{g^4}{16\pi} F_2(s,M_\chi),\nn\\
&& a=g_{CG}\frac{g^4}{32 \pi  M_{\chi }^2},\quad b=g_{CG}\frac{3 g^4}{256 \pi  M_{\chi }^2}.
\eea
\item $\chi_m \bar{\chi}_{m-2} \to W^+ W^+$~:
\bea
&& g_{\rm CG}=(j+m)(j+m-1)(j-m+1)(j-m+2),\nn\\
&&\sigma =g_{\rm CG}\frac{g^4}{32\pi} F_2(s,M_\chi),\nn\\
&& a=g_{CG}\frac{g^4}{64 \pi  M_{\chi }^2},\quad b=g_{CG}\frac{3 g^4}{512 \pi  M_{\chi }^2};
\eea
\item $\chi_m\bar{\chi}_m\to ZZ$~:
\bea
&& g_{\rm CG}=m^4,\nn\\
&& \sigma=g_{CG}\frac{\left(e^2-g^2\right)^2}{8 \pi }F_2(s,M_\chi)\nn\\
&&a=g_{CG}\frac{\left(e^2-g^2\right)^2}{16 \pi  M_{\chi }^2}, \quad b=g_{CG}\frac{3 \left(e^2-g^2\right)^2}{128 \pi  M_{\chi }^2};
\eea
\item $\chi_m\bar{\chi}_m\to \gamma\gamma$~:
\bea
&& g_{\rm CG}=m^4,\nn\\
&& \sigma=g_{CG}\frac{e^4}{8 \pi }F_2(s,M_\chi),\nn\\
&&a=g_{CG}\frac{e^4}{16 \pi  M_{\chi }^2}, \quad b=g_{CG}\frac{3 e^4}{128 \pi  M_{\chi }^2};
\eea
\item $\chi_m\bar{\chi}_m\to \gamma Z$~:
\bea
&& g_{\rm CG}=m^4,\nn\\
&& \sigma=g_{CG}\frac{e^2(g^2-e^2)}{4 \pi }F_2(s,M_\chi),\nn\\
&&a=g_{CG}\frac{e^2(g^2-e^2)}{8 \pi  M_{\chi }^2}, \quad b=g_{CG}\frac{3e^2(g^2-e^2)}{64 \pi  M_{\chi }^2}.
\eea
\end{itemize}

Now consider the processes involving more complicated kinematics:
\begin{itemize}[leftmargin=*]
\item $\chi_0\bar{\chi}_{-1}(\chi_1\bar{\chi}_0)\to \gamma W^+$~:
\bea
&& g_{CG}=j(j+1),\quad j\geqslant1\nn\\
&&\sigma=g_{CG}\frac{e^2g^2}{48\pi}F_3(s,M_\chi),\nn\\
&& a = g_{CG}\frac{e^2g^2}{64\pi M_\chi^2},\quad b = g_{CG}\frac{11e^2g^2}{1536\pi M_\chi^2};
\eea
\item $\chi_{-1}\bar{\chi}_{-2}(\chi_2\bar{\chi}_{1})\to \gamma W^+$~:
\bea
&& g_{CG}=(j-1)(j+2),\quad j\geqslant2,\nn\\
&&\sigma=g_{CG}\frac{e^2g^2}{48\pi}F_4(s,M_\chi),\nn\\
&& a = g_{CG}\frac{9 e^2g^2}{64\pi M_\chi^2},\quad b = g_{CG}\frac{83e^2g^2}{1536\pi M_\chi^2};
\eea
\item $\chi_{-2}\bar{\chi}_{-3}(\chi_3\bar{\chi}_{2})\to \gamma W^+$~:
\bea
&& g_{CG}=(j-2)(j+3),\quad j\geqslant3,\nn\\
&&\sigma=g_{CG}\frac{e^2g^2}{48\pi}F_5(s,M_\chi),\nn\\
&& a = g_{CG}\frac{75 e^2g^2}{192\pi M_\chi^2},~ b = g_{CG}\frac{227e^2g^2}{1536\pi M_\chi^2};
\eea
\item $\chi_0 \bar{\chi}_{-1}(\chi_1 \bar{\chi}_0) \to Z W^+$~:
\bea
&& g_{CG}=j(j+1),\quad j\geqslant1, \nn\\
&&\sigma=g_{CG}\frac{g^2}{384\pi}F_6(s,M_\chi),\nn\\
&& a = g_{CG}\frac{9g^4-8e^2g^2}{512\pi M_\chi^2},~b = g_{CG}\frac{83g^4-88e^2g^2}{12288\pi M_\chi^2};
\eea
\item $\chi_{-1} \bar{\chi}_{-2}(\chi_2 \bar{\chi}_1) \to Z W^+$~:
\bea
&& g_{CG}=(j-1)(j+2),\quad j\geqslant2, \nn\\
&&\sigma=g_{CG}\frac{g^2}{384\pi}F_7(s,M_\chi),\nn\\
&& a = g_{CG}\frac{73g^4-72e^2g^2}{512\pi M_\chi^2},\nn\\
&&b = g_{CG}\frac{659g^4-664e^2g^2}{12288\pi M_\chi^2};
\eea
\item $\chi_{-2} \bar{\chi}_{-3}(\chi_3 \bar{\chi}_2) \to Z W^+$~:
\bea
&& g_{CG}=(j-2)(j+3),\quad j\geqslant2, \nn\\
&&\sigma=g_{CG}\frac{g^2}{384\pi}F_8(s,M_\chi),\nn\\
&& a = g_{CG}\frac{603g^4-600e^2g^2}{1536\pi M_\chi^2},\nn\\
&&b = g_{CG}\frac{1811g^4-1816e^2g^2}{12288\pi M_\chi^2};
\eea
\end{itemize}

Finally, we present the  results of  the most complicated channel which consists of $W^+W^-$ boson pairs in the final state. The results read as follows:
\begin{itemize}[leftmargin=*]
\item $\chi_{j} \bar{\chi}_{j} \to W^+ W^-$ for $j=1,2,3$~:
\bea
&& g_{CG}=j^2,\nn\\
&&\sigma=g_{CG}\frac{g^4}{64\pi}F_9(s,M_\chi),\nn\\
&&a = g_{CG}\frac{9g^4}{256\pi M_\chi^2},\quad b = g_{CG}\frac{3g^4}{64\pi M_\chi^2};
\eea
\item $\chi_{1} \bar{\chi}_{1} \to W^+ W^-$ for the D2 model ($j=2$)~:
\bea
&&\sigma=\frac{g^4}{64\pi}F_{10}(s,M_\chi),\nn\\
&&a = \frac{201g^4}{256\pi M_\chi^2},\quad b = \frac{1811g^4}{6144\pi M_\chi^2};
\eea
\item $\chi_{1} \bar{\chi}_{1} \to W^+ W^-$ for the D3 model ($j=3$)~:
\bea
&&\sigma=\frac{g^4}{64\pi}F_{11}(s,M_\chi),\nn\\
&&a = \frac{969g^4}{256\pi M_\chi^2},\quad b = \frac{8723g^4}{6144\pi M_\chi^2};
\eea
\item $\chi_{2} \bar{\chi}_{2} \to W^+ W^-$ for the D3 model ($j=3$)~:
\bea
&&\sigma=\frac{g^4}{16\pi}F_{12}(s,M_\chi),\nn\\
&&a = \frac{129g^4}{64\pi M_\chi^2},\quad b = \frac{1163g^4}{1536\pi M_\chi^2};
\eea
\end{itemize}

\subsection{Gauge boson plus Higgs boson}

As the Higgs boson does not couple to the dark particles, the annihilation mode of a gauge boson in association with a Higgs boson occur through a $s$-channel diagram. The cross sections and coefficients are summarized as follows:
\begin{itemize}[leftmargin=*]
\item $\chi_m\bar{\chi}_{m-1} \to W^*\to W^+H$~:
\bea
&& g_{CG}=(j+m)(j-m+1),\nn\\
&&\sigma=g_{CG}\frac{g^4}{384\pi} F_1(s,M_\chi),\nn\\
&& a = g_{CG}\frac{g^4}{512\pi M_\chi^2},\quad b = -g_{CG}\frac{5g^4}{12288\pi M_\chi^2};
\eea
\item $\chi_m\bar{\chi}_{m} \to Z^*\to ZH$~:
\bea
&& g_{CG}=m^2,\nn\\
&&\sigma=g_{CG}\frac{g^4}{192\pi} F_1(s,M_\chi),\nn\\
&& a = g_{CG}\frac{g^4}{256\pi M_\chi^2},\quad b = -g_{CG}\frac{5g^4}{6144\pi M_\chi^2}.
\eea
\end{itemize}

\section{Annihilation cross sections of Majorana-type DM }\label{sec:xsection-Majorana}

We can easily obtain the cross sections and coefficients of the Majorana-type DMs from those of the Dirac-type DMs. The only difference is the numbers of annihilation channels of the Majorana DMs is one half of the numbers of the Dirac DMs. We list out the annihilation channels of Majorana DMs and the corresponding cross sections and coefficients can be read out from those of Dirac DMs given in Sec.~\ref{sec:xsection-Dirac}.

\bibliographystyle{apsrev}
\bibliography{reference}

\end{document}